\documentclass[superscriptaddress,twocolumn,aps,preprintnumbers,notitlepage,groupedaddress]{revtex4-1}
\usepackage{amsmath,amssymb,esint}
\usepackage{lipsum} 
\usepackage[latin9]{inputenc}
\usepackage{graphicx}
\usepackage{dcolumn}
\usepackage{bbold}
\usepackage{bm}
\usepackage[usenames,table]{xcolor}
\usepackage[colorlinks=true,linktocpage,bookmarks=true,citecolor=blue,linkcolor=blue,urlcolor=blue,hyperfootnotes=true,pageanchor=true]{hyperref}
\usepackage[caption=false,labelformat=simple]{subfig}
\usepackage{accents}
\usepackage[export]{adjustbox}
\usepackage{mathtools}
\usepackage{float}
\usepackage{mathrsfs}
\usepackage{pifont}

\usepackage{adjustbox}
\usepackage{multirow}

\usepackage{soul} 

\begin{document}

\title{Unconventional superconductivity in altermagnets with spin-orbit coupling}

\author{Vanuildo S. de Carvalho}
\affiliation{Instituto de F\'{\i}sica, Universidade Federal de Goi\'as, 74.690-900, Goi\^ania-GO,
Brazil}
\author{Hermann Freire}
\affiliation{Instituto de F\'{\i}sica, Universidade Federal de Goi\'as, 74.690-900, Goi\^ania-GO,
Brazil}

\begin{abstract}
We investigate some possible symmetries of the superconducting state that emerges in three-dimensional altermagnets in the presence of spin-orbit coupling. We demonstrate within a weak-coupling approach that these altermagnets, which naturally possess an order modulated by a vector form factor $\boldsymbol{g}_{\mathbf{k}}$, favor spin-triplet superconductivity described by gap functions given by $\boldsymbol{d}(\mathbf{k}) = \boldsymbol{u}(\mathbf{k}) \times \boldsymbol{g}_{\mathbf{k}}$, where $\boldsymbol{u}(\mathbf{k}) = - \boldsymbol{u}(-\mathbf{k})$. Consequently, this singles out $f$-wave spin-triplet superconductivity as the most favorable pairing state to appear in the vicinity of $d$-wave altermagnetism. Furthermore, we obtain that the combination of spin-singlet superconducting states with altermagnetism gives rise to Bogoliubov-Fermi surfaces, which are protected by a $\mathbb{Z}_2$ topological invariant. Using a Ginzburg-Landau analysis, we show that, for a class of spin-orbit coupled altermagnetic models, a superconducting phase is expected to appear at low temperatures as an intertwined $d + if$ state, thus breaking time-reversal symmetry spontaneously.
\end{abstract}

\date{\today}

\maketitle

\emph{Introduction.} Altermagnetism is an intriguing magnetic phase  discovered recently \cite{Smejkal2022_1,Smejkal2022_2,Jungwirth-arXiv(2024)} that intertwines with the properties of both ferromagnetism and antiferromagnetism. In altermagnets, the combination of time-reversal and rotation symmetries leads to zero bulk magnetization and to a non-uniform Zeeman splitting of the band structure, which becomes nodal at symmetry-related points of the reciprocal space. This new magnetic phase represents a departure from the usual dichotomy that was previously assumed to exist for collinear magnets and, for this reason, has attracted widespread attention of the condensed matter physics community in the last few years. In this regard, RuO$_2$ \cite{Fedchenko_2024} and MnTe \cite{Libor_MnTe} were the first experimentally confirmed compounds to exhibit altermagnetism. In addition, other potential candidate materials consist of, e.g., the parent cuprate compound La$_2$CuO$_4$ \cite{Smejkal2022_1}, the semimetal CoNb$_3$S$_6$ \cite{Smejkal2022_3}, the pnictide with metal-insulator transition FeSb$_2$ \cite{Mazin_PNAS}, among many others. 

Although this phase was proposed in the context of new applications in the field of spintronics, another important aspect of it lies in its possible relation to the field of
superconductivity \cite{Mazin2022notes}. Naturally, superconductivity and magnetism are two fundamental phenomena that are usually intertwined to each other: it has long been pointed out that magnetic fluctuations generally promote unconventional superconductivity in many important quantum materials \cite{Sigrist-RMP(1991)}. For instance, antiferromagnetically mediated interactions usually lead to the formation spin-singlet Cooper pairs \cite{Scalapino_1986,Monod_Emery,Miyake_Varma}, while ferromagnetic fluctuations are in turn responsible for the emergence of spin-triplet superconductivity \cite{Fay_Layzer,Berk_Schrieffer}. Since altermagnetism stands out as a new magnetic phase with simultaneous features of both antiferromagnetism and ferromagnetism, it could in principle drive the formation of either spin-singlet or spin-triplet superconductivity, or even lead to a more exotic scenario of displaying a mixture of both superconducting (SC) phases \cite{Mazin2022notes}.

The investigation of microscopic mechanisms of possible intrinsic SC phases in altermagnetic (AM) models has initiated very recently, soon after the discovery of this magnetic phase \cite{Smejkal2022_1,Smejkal2022_2}. Although altermagnetism can appear in both two- and three-dimensional (2D and 3D) systems \cite{Schiff-arXiv(2024),McClarty-PRL(2024)}, most works to date have focused on the analysis of the pairing properties of effective 2D AM models \cite{Ouassou-PRL(2023),Papaj2023,Neupert2024,Zhu2023,Sudbo2023,Chakraborty2024,Hong2024,Chakraborty-arXiv(2024),Scheurer-PRB(2024),Knolle-arXiv(2024)}. Recently, it has been observed that the 3D compound CrSb turns out to be a topological Weyl altermagnet \cite{vanderBrink2024,Ma-arXiv(2024)}. Interestingly, the appearance of Weyl nodes in altermagnets was predicted to exist when the spin-orbit coupling (SOC) becomes relevant in these systems \cite{Fernandes-PRB(2024),Venderbos-arXiv(2024)}. 

Motivated in part by these experimental findings, we study in this work the pairing instabilities of spin-orbit coupled AM models in 3D \cite{Fernandes-PRB(2024)}. We find that the AM order modulated by a vector form factor $\boldsymbol{g}_{\mathbf{k}}$ is beneficial to spin-triplet SC states having gap functions $\boldsymbol{d}(\mathbf{k}) = \boldsymbol{u}(\mathbf{k}) \times \boldsymbol{g}_{\mathbf{k}}$, where $\boldsymbol{u}(\mathbf{k}) = - \boldsymbol{u}(-\mathbf{k})$ due to Fermi-Dirac statistics. As a result, if $\boldsymbol{g}_{\mathbf{k}}$ has $d$-wave symmetry, the corresponding AM interaction will favor $f$-wave spin-triplet SC as the most favorable pairing state at low temperatures. Importantly, the combination of spin-singlet SC and altermagnetism leads the system to exhibit Bogoliubov-Fermi surfaces \cite{Agterberg-PRL(2017),Brydon-PRB(2018)}, which become topologically protected due to the presence of a $\mathbb{Z}_2$ invariant. We also show that, for a class of AM models, a SC phase is expected to appear displaying $d + if$ symmetry, which possesses both singlet and triplet amplitudes and breaks spontaneously time-reversal symmetry.

\begin{figure}[t]
\centering
\centering \includegraphics[width=0.9\linewidth,valign=t]{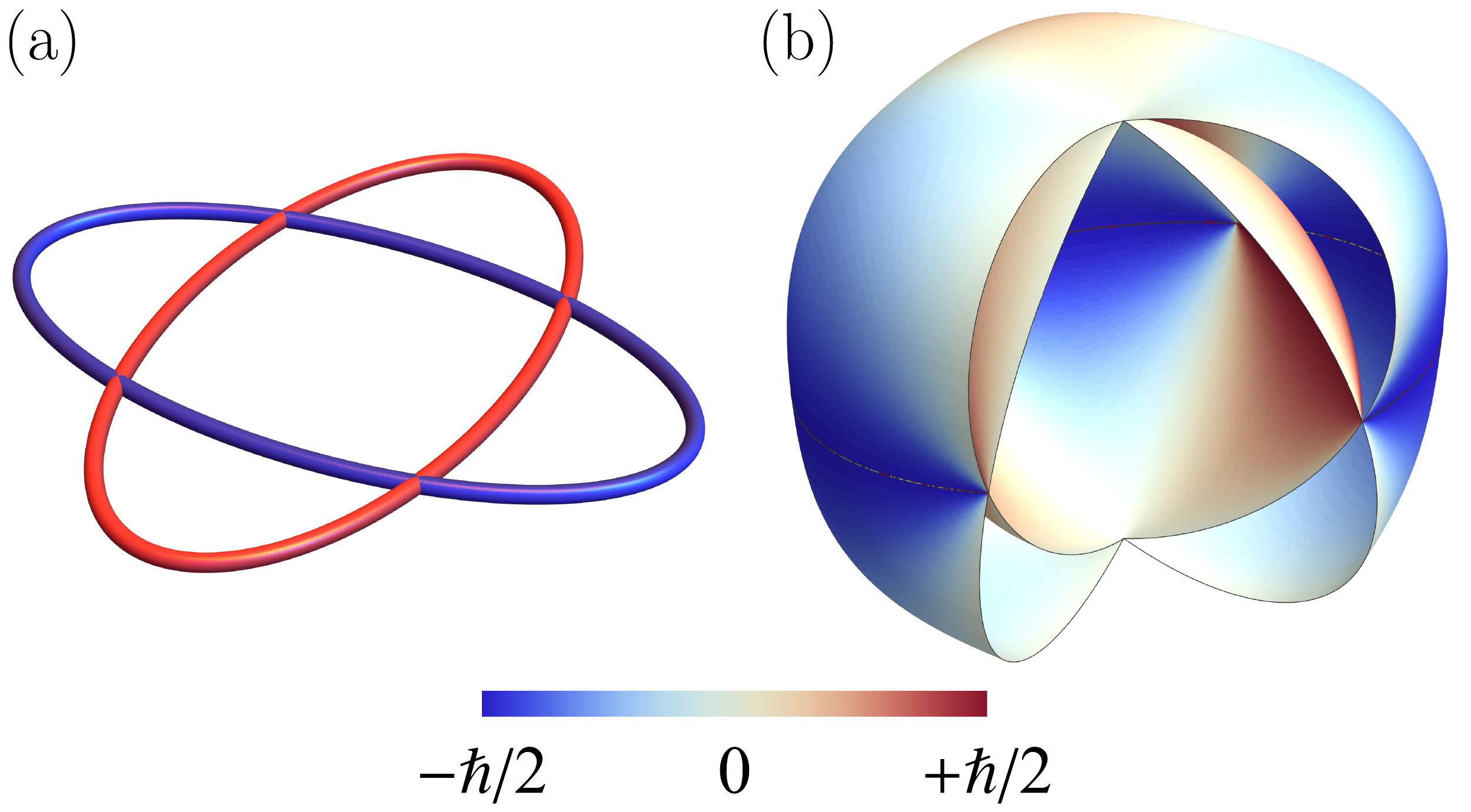}
\caption{Schematic representation of the Fermi surfaces for altermagnets (a) without SOC and (b) with SOC. The former system is effectively 2D, whereas the latter displays 3D behavior. The scale at the bottom indicates the spin texture of the Fermi states. While in 2D altermagnets, the spin texture only assumes one of the values $\pm \hbar/2$, it varies continuously between $- \hbar/2$ and $+ \hbar/2$ for 3D systems. In addition, the latter realize type-II Weyl nodes at the pinching points of their Fermi surface sheets. The interior of the Fermi surface in (b) is exposed for visualization purposes.}\label{Spin_Texture}
\end{figure}

\emph{Model.} We consider an AM metal in which the non-interacting Hamiltonian reads $H_0 = \sum_{\mathbf{k}, s} \xi_{\mathbf{k}} \psi^{\dagger}_{\mathbf{k}, s} \psi_{\mathbf{k}, s}$,  where $\psi^{\dagger}_{\mathbf{k}, s}$ ($\psi_{\mathbf{k}, s}$) represents the creation (annihilation) operators for electrons with a parabolic band dispersion $\xi_{\mathbf{k}} = \frac{\mathbf{k}^2}{2 m} - \mu$, spin projection $s \in \{ \uparrow, \downarrow\}$, and chemical potential $\mu$. As shown recently \cite{Fernandes-PRB(2024)}, if one further assumes that the SOC connecting the lattice and the spin degrees of freedom is significant, the components of the local spin-moment magnetization turn out to be independent from each other. As a result, for systems in which the AM order parameter transforms as a one-dimensional irreducible representation (irrep) of the point group, the coupling between AM and the electronic degrees of freedom becomes described by:
\begin{equation}
H_{\mathrm{AM}} = \lambda \sum_{\mathbf{k}, s, s'}\psi^{\dagger}_{\mathbf{k}, s}[\boldsymbol{g}_{\mathbf{k}} \cdot \hat{\boldsymbol{\sigma}}_{s, s'} ]\psi^{\dagger}_{\mathbf{k}, s'}.
\end{equation} 
Here, $\lambda$ refers to the AM interaction, $\hat{\boldsymbol{\sigma}} = (\hat{\sigma}_x, \hat{\sigma}_y, \hat{\sigma}_z)$ denotes the vector of Pauli matrices, and $\boldsymbol{g}_{\mathbf{k}}$ is the form factor, which depends on the irrep of the AM order parameter (see Ref. \cite{Fernandes-PRB(2024)}). For example, when the AM system transforms as the $B^{-}_{1g}$ irrep of the tetragonal group $D_{4h}$, one has $\boldsymbol{g}_{\mathbf{k}} = k_y k_y \hat{\boldsymbol{x}} + k_x k_z \hat{\boldsymbol{y}} + k_x k_y \hat{\boldsymbol{z}}$ for the momentum in the vicinity of the $\Gamma$ point. The independent components of $\boldsymbol{g}_{\mathbf{k}}$ are the main consequence of the SOC interaction; they lead to non-collinear AM states and to anisotropic spin textures of the band structure (see Fig. \ref{Spin_Texture}). The Hamiltonian $H_{\mathrm{AM}}$ is characterized by a set of nodal lines defined by $\boldsymbol{g}_{\mathbf{k}} = \boldsymbol{0}$. It is also odd under time-reversal symmetry and exhibits charge-conjugation symmetry: $\widehat{C}^{-1} \widehat{\mathcal{H}}_{\mathrm{AM}}(\mathbf{k}) \widehat{C} = - \widehat{\mathcal{H}}_{\mathrm{AM}}(\mathbf{k})$, where $\widehat{C} = \mathcal{K} \hat{\sigma}_{y}$ and $\mathcal{K}$ denotes complex conjugation. Because $\widehat{C}^2 = - \mathbb{1}$, $\widehat{\mathcal{H}}_{\mathrm{AM}}$ belongs to the class C of the periodic table of gapless topological phases \cite{Schnyder2014,Chiu2015}, which means that $H_{\mathrm{AM}}$ is trivial under non-spatial symmetries. On the other hand, because these nodal lines lie on mirror planes of the point group, the theory of gapless topological phases also predicts that they are protected by reflection symmetries. Besides, the AM system $H_0 + H_{\mathrm{AM}}$ displays Fermi-surface pinching points behaving as type-II Weyl nodes \cite{Bernevig2015,Zhang-PRL(2015)}, which are localized at the crossing points between the nodal lines and the non-interacting Fermi surface.

As mentioned earlier, several works have recently addressed the interplay between altermagnetism and superconductivity in 2D systems \cite{Ouassou-PRL(2023),Papaj2023,Neupert2024,Zhu2023,Sudbo2023,Chakraborty2024,Hong2024,Chakraborty-arXiv(2024)}. However, those studies assume that SOC is negligible in the parent AM phase, which ultimately prevents the appearance of nodal lines and Weyl nodes in the band structure. Since symmetry-protected nodal lines are also expected to affect other properties of altermagnets \cite{Fernandes-PRB(2024),Venderbos-arXiv(2024)}, henceforth we investigate their impact on the formation of a SC state. For completeness, we will consider here an effective model possessing an attractive interaction $V_{\mathbf{k}, \mathbf{k}'}$, which equally favors both spin-singlet and spin-triplet Cooper pairings. Thus, the total Hamiltonian becomes described by $H = H_0 + H_{\mathrm{AM}} + H_{\mathrm{SC}}$, where the term
\begin{equation}
\hspace{-0.3cm} H_{\mathrm{SC}} \hspace{-0.05cm} = \hspace{-0.05cm} \frac{1}{2 \mathcal{V}} \hspace{-0.1cm} \sum_{\substack{\mathbf{k}, \mathbf{k}', \mathbf{q} \\ s, s'}} \hspace{-0.15cm} V_{\mathbf{k}, \mathbf{k}'} \psi^{\dagger}_{\mathbf{k} + \frac{\mathbf{q}}{2}, s} \psi^{\dagger}_{- \mathbf{k} + \frac{\mathbf{q}}{2}, s'} \psi_{- \mathbf{k}' + \frac{\mathbf{q}}{2}, s'} \psi_{\mathbf{k}' + \frac{\mathbf{q}}{2}, s}\hspace{-0.2cm}\label{Eq_SC_Ham}
\end{equation}
drives the SC instability and $\mathcal{V}$ denotes the volume of the system.

\begin{figure*}[t]
\centering
\centering \includegraphics[width=0.45\linewidth,valign=t]{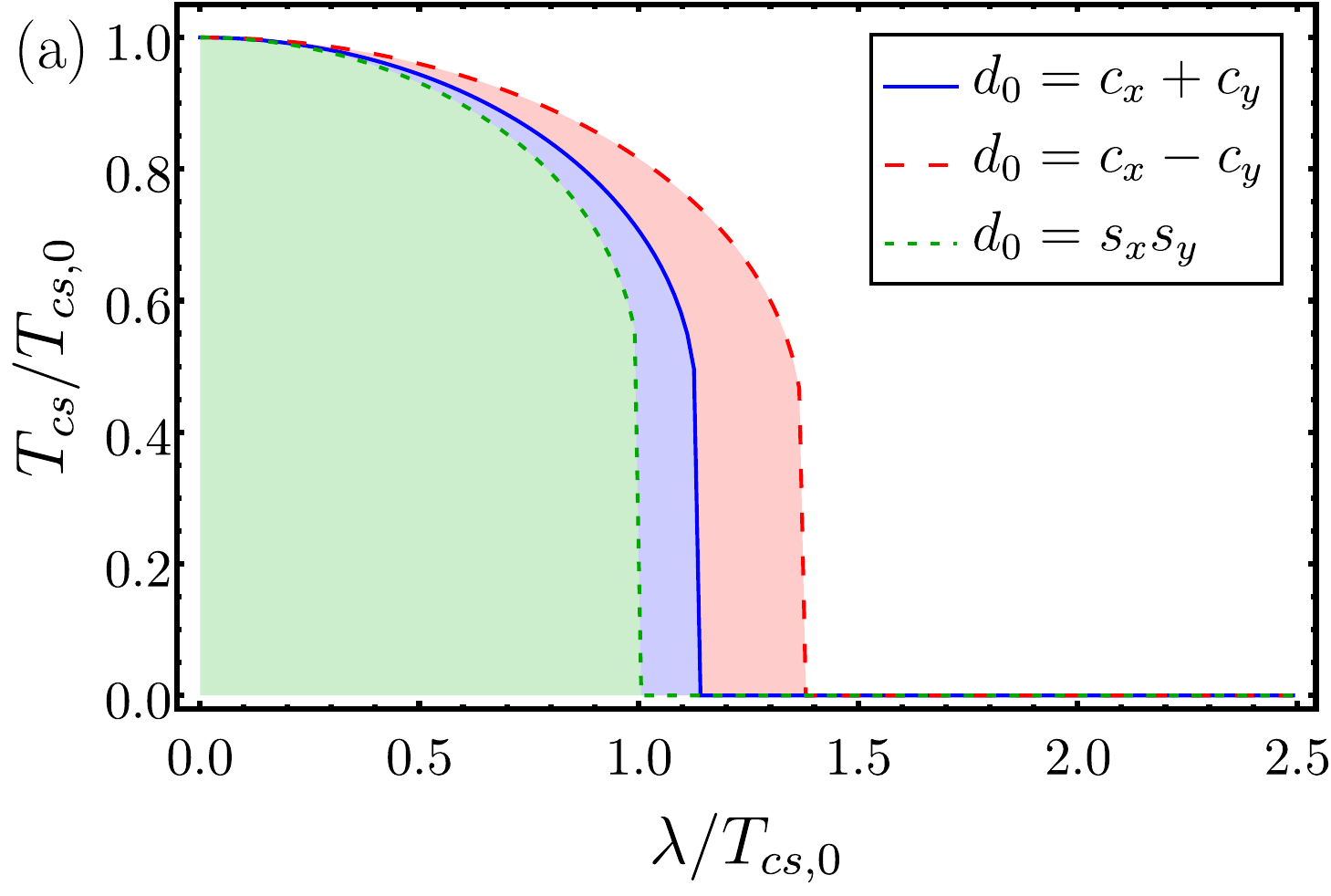} \hfil{} \includegraphics[width=0.45\linewidth,valign=t]{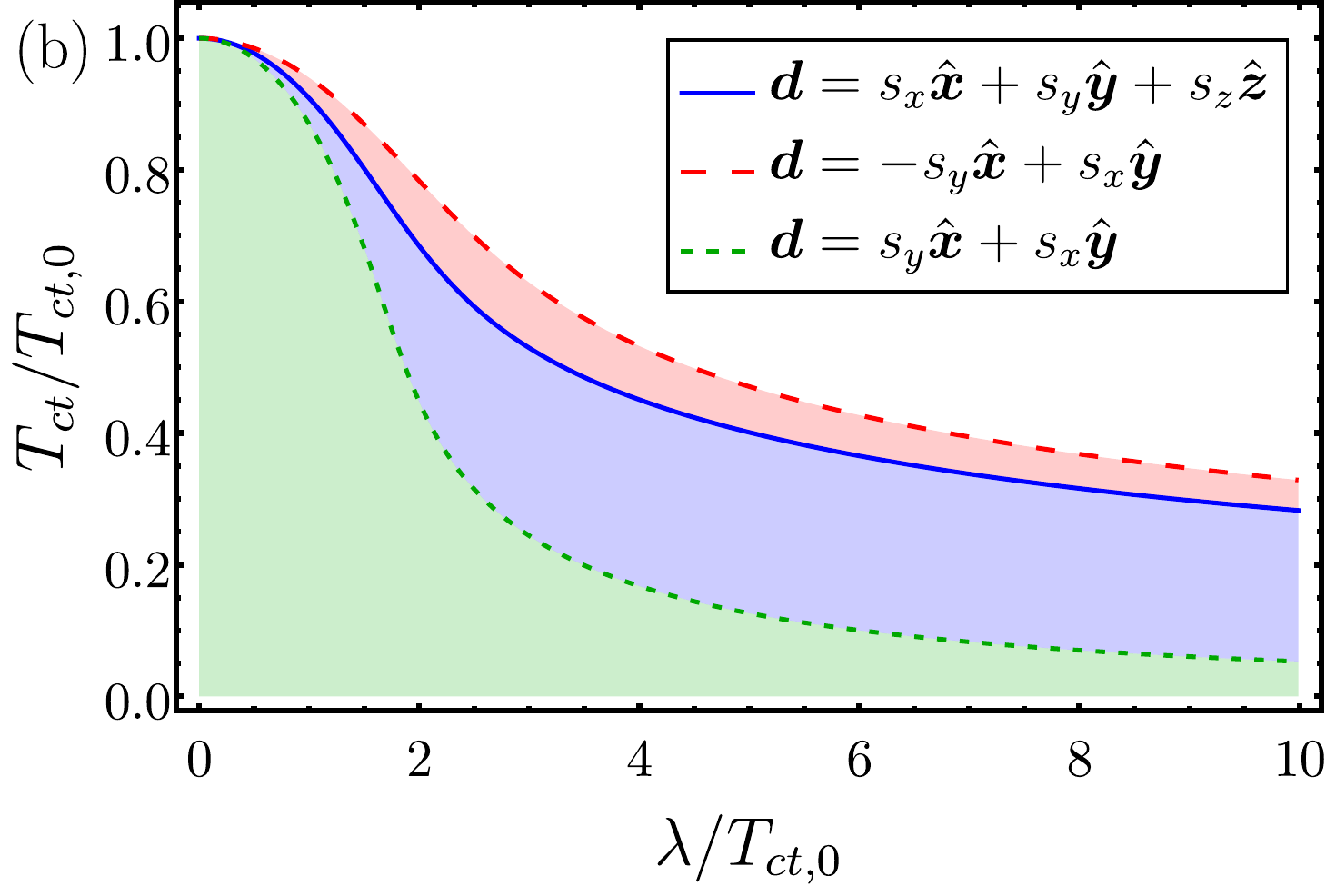}
\caption{Impact of the interaction $\lambda$ due to a $B^{-}_{1g}/D_{4h}$ AM phase on the critical temperatures $T_{cs}$ and $T_{ct}$ for SC phases in the (a) spin-singlet and (b) spin-triplet channels, respectively. The gap functions $d_0(\mathbf{k})$ and $\boldsymbol{d}(\mathbf{k})$ are written in terms of $c_a = \cos k_a$ and $s_a = \sin k_a$ with $a = x,y,z$. In (a), $d_0(\mathbf{k})$ has (from top to bottom) extended $s$-wave, $d_{x^2 - y^2}$-wave and $d_{xy}$-wave symmetries, while in (b) $\boldsymbol{d}(\mathbf{k})$ represents SC states with $p$-wave symmetry. The increase of $\lambda$ is always detrimental to both $T_{cs}$ and $T_{ct}$ for the SC orders shown in the above plots; however, only $T_{cs}$ is completely suppressed for finite $\lambda$.}\label{Tc_Altermag}
\end{figure*}

\emph{SC instabilities of 3D altermagnets with SOC.} We investigate now the formation of the SC state within a weak-coupling BCS theory \cite{Sigrist-RMP(1991),Coleman-CUP(2015)}: $V_{\mathbf{k}, \mathbf{k}'}$ is finite within a cutoff energy $\epsilon_c$ around the Fermi surface and depends on the momentum through the angular coordinates. After using a mean-field decoupling for $H_{\mathrm{SC}}$ [see Supplemental Material (SM) \cite{Suppl_Mat} for details], we find that the Bogoliubov-de Gennes (BdG) Hamiltonian related to $H = H_0 + H_{\mathrm{AM}} + H_{\mathrm{SC}}$ becomes:
\begin{equation}
H_{\mathrm{BdG}} = \frac{1}{2} \sum_{\mathbf{k}} \Psi^{\dagger}_{\mathbf{k}} \begin{pmatrix} \widehat{\mathcal{H}}_0(\mathbf{k}) & \widehat{\Delta}(\mathbf{k}) \\ \widehat{\Delta}^{\dagger}(\mathbf{k}) & - \hat{\sigma}_y \widehat{\mathcal{H}}^{T}_0(- \mathbf{k}) \hat{\sigma}_y \end{pmatrix} \Psi_{\mathbf{k}},
\end{equation}
where $\Psi_{\mathbf{k}} \equiv (\psi_{\mathbf{k}, \uparrow}, \; \psi_{\mathbf{k}, \downarrow}, \; - \psi^{\dagger}_{-\mathbf{k}, \downarrow}, \; \psi^{\dagger}_{-\mathbf{k}, \uparrow})^T$ is the Balian-Werthamer spinor and $\widehat{\mathcal{H}}_0(\mathbf{k}) \equiv \xi_{\mathbf{k}} \hat{\sigma}_0 + \lambda \boldsymbol{g}_{\mathbf{k}} \cdot \hat{\boldsymbol{\sigma}}$ is the lattice Hamiltonian for a pure AM phase. In addition, $\widehat{\Delta}(\mathbf{k})$ is the SC order parameter which is allowed to possess both spin-singlet and spin-triplet components. Hence, it is parametrized as $\widehat{\Delta}(\mathbf{k}) = \Delta_s(\mathbf{k}) \hat{\sigma}_0 + \boldsymbol{\Delta}_t(\mathbf{k}) \cdot \hat{\boldsymbol{\sigma}}$, where $\Delta_s(\mathbf{k}) = \vert \Delta_s \vert e^{i \theta_s} d_0(\mathbf{k})$ and $\boldsymbol{\Delta}_t(\mathbf{k}) = \vert \Delta_t \vert e^{i \theta_t} \boldsymbol{d}(\mathbf{k})$ refer to the singlet and triplet pairing amplitudes, respectively. Notice that the SC state breaks time-reversal symmetry provided the phase difference $\theta_s - \theta_t$ is non-trivial. Besides, as required by Fermi-Dirac statistics, $d_0(\mathbf{k}) = d_0(- \mathbf{k})$ and $\boldsymbol{d}(\mathbf{k}) = - \boldsymbol{d}(- \mathbf{k})$ \cite{Sigrist-RMP(1991)}.

To determine the equation for the SC order parameter, we minimize the effective SC action $\mathcal{S}_{\mathrm{eff}}[\widehat{\Delta}^\dagger, \widehat{\Delta}]$ related to $H_{\mathrm{BdG}}$ with respect to $\widehat{\Delta}$. Assuming that $\widehat{\Delta}(\mathbf{k})$ has the same dependence on the two AM Fermi surfaces, this approach yields:
\begin{align}\label{Eq_SC_OP01}
\widehat{\Delta}^{\dagger}(\mathbf{k}) & = \frac{1}{\beta \mathcal{V}} \sum_{n, \mathbf{k}'} V_{\mathbf{k}, \mathbf{k}'}\{ [\widehat{G}_{0, h}(\mathbf{k}', i\omega_n)]^{-1} - \widehat{\Delta}^{\dagger}(\mathbf{k}') \nonumber \\
& \times  \widehat{G}_{0, p}(\mathbf{k}', i\omega_n)\widehat{\Delta}(\mathbf{k}') \}^{-1} \widehat{\Delta}^{\dagger}(\mathbf{k}') \widehat{G}_{0, p}(\mathbf{k}', i\omega_n).
\end{align}
$\widehat{G}_{0, p}(\mathbf{k}, i \omega_n)$ and $\widehat{G}_{0, h}(\mathbf{k}, i \omega_n)$ are, respectively, the particle and hole propagators. To calculate the transition temperature $T_c$, we then linearize Eq. \eqref{Eq_SC_OP01} with respect to $\widehat{\Delta}(\mathbf{k})$, which leads to the following set of singlet and triplet gap equations: 
\begin{align}
\Delta^{*}_s(\mathbf{k}) & = - \frac{1}{\beta \mathcal{V}} \sum_{n, \mathbf{k}'} V_{\mathbf{k}, \mathbf{k}'} [ (\widetilde{G}_{+}G_{+} - \widetilde{G}_{-}G_{-}) \Delta^{*}_s(\mathbf{k}') \nonumber \\
& + (\widetilde{G}_{+}G_{-} - \widetilde{G}_{-}G_{+}) \boldsymbol{\Delta}^{*}_t(\mathbf{k}') \cdot \widehat{\boldsymbol{g}}_{\mathbf{k}'} ], \label{Eq_SC_OP02}\\
\boldsymbol{\Delta}^{*}_t(\mathbf{k}) & = - \frac{1}{\beta \mathcal{V}} \sum_{n, \mathbf{k}'} V_{\mathbf{k}, \mathbf{k}'} \lbrace (\widetilde{G}_{+} G_{+} - \widetilde{G}_{-} G_{-}) \boldsymbol{\Delta}^{*}_t(\mathbf{k}') \nonumber \\
& - 2 \widetilde{G}_{-} G_{-} [(\boldsymbol{\Delta}^{*}_t(\mathbf{k}') \cdot \widehat{\boldsymbol{g}}_{\mathbf{k}'}) \widehat{\boldsymbol{g}}_{\mathbf{k}'} - \boldsymbol{\Delta}^{*}_t(\mathbf{k}')] \nonumber \\
& + (\widetilde{G}_{+} G_{-} - \widetilde{G}_{-} G_{+})\Delta^{*}_s(\mathbf{k}') \widehat{\boldsymbol{g}}_{\mathbf{k}'} + i (\widetilde{G}_{+} G_{-} \nonumber \\
& + \widetilde{G}_{-} G_{+}) \boldsymbol{\Delta}^{*}_t(\mathbf{k}') \times \widehat{\boldsymbol{g}}_{\mathbf{k}'} \rbrace, \label{Eq_SC_OP03}
\end{align}
where $\widetilde{G}_{a} G_{b} \equiv G_{a}(- \mathbf{k}, - i \omega_n) G_{b}(\mathbf{k}, i \omega_n)$, $G_{\pm}(\mathbf{k}, i \omega_n) = \frac{1}{2}\left[ (-i \omega_n + E_{\mathbf{k}, +})^{-1} \pm (-i \omega_n + E_{\mathbf{k}, -})^{-1} \right]$, $E_{\mathbf{k}, \pm} \equiv \xi_{\mathbf{k}} \pm \lambda \| \boldsymbol{g}_{\mathbf{k}} \|$, and $\widehat{\boldsymbol{g}}_{\mathbf{k}} \equiv \boldsymbol{g}_{\mathbf{k}}/\| \boldsymbol{g}_{\mathbf{k}}\|$.  

The spin-singlet and spin-triplet gap equations [Eqs. \eqref{Eq_SC_OP02} and \eqref{Eq_SC_OP03}] become coupled when the AM interaction $\lambda$ is finite. As occurs in the BCS theory of noncentrosymmetric superconductors \cite{Sigrist-PRL(2004)}, this coupling depends on the particle-hole asymmetry of the model, which is a measure of the difference of the density of states $N(E)$ on the two AM Fermi surfaces. For simplicity, we will approximate $N(E)$ by its value at the Fermi level ($N_F$) for both Fermi surfaces, which will thus decouple Eqs. \eqref{Eq_SC_OP02} and \eqref{Eq_SC_OP03} from each other. The latter procedure is allowed within a weak-coupling theory, because the most important contributions are dominated by fermionic states in the vicinity of the corresponding Fermi surfaces. Additionally, we also consider that the gap functions and the pairing interaction $V_{\mathbf{k}, \mathbf{k}'}$ are normalized over the Fermi surfaces as $\langle \vert \Delta_s(\mathbf{k}) \vert^2 \rangle_{\mathbf{k}} = \vert \Delta_s \vert^2$, $\langle \| \boldsymbol{\Delta}_t(\mathbf{k}) \|^2 \rangle_{\mathbf{k}} = \vert \Delta_t \vert^2$, $\langle V_{\mathbf{k}, \mathbf{k}'} \Delta_s(\mathbf{k}) \rangle_{\mathbf{k}} = - V_s \Delta_s(\mathbf{k}')$, $\langle V_{\mathbf{k}, \mathbf{k}'} \boldsymbol{\Delta}_t(\mathbf{k}) \rangle_{\mathbf{k}} = - V_t \boldsymbol{\Delta}_t(\mathbf{k}')$, where $\langle (\cdots) \rangle_{\mathbf{k}} \equiv \int_{\mathcal{S}^2} \frac{d\Omega_{\mathbf{k}}}{4 \pi} (\cdots)$ with $\Omega_{\mathbf{k}}$ being the solid angle, and $V_s$ ($V_t$) is the pairing coupling for the spin-singlet (spin-triplet) SC state. As a result, we obtain the gap equations: 
\begin{align}
\frac{\vert \Delta_s \vert^2}{N_F V_s} & = \int_{\mathcal{S}^2} \frac{d\Omega_{\mathbf{k}}}{4\pi} \left[ \ln\left(\frac{2 e^{\gamma} \epsilon_c}{\pi T_{cs}} \right) -\Upsilon\left(\frac{\lambda \| \boldsymbol{g}_{\mathbf{k}} \|}{2 \pi T_{cs}} \right) \right] \vert \Delta_s(\mathbf{k}) \vert^2, \label{Eq_SC_OP04} \\
\frac{\vert \Delta_t \vert^2}{N_F V_t} & = \int_{\mathcal{S}^2} \frac{d\Omega_{\mathbf{k}}}{4\pi} \bigg[ \ln\left(\frac{2 e^{\gamma} \epsilon_c}{\pi T_{ct}} \right) \| \boldsymbol{\Delta}_t(\mathbf{k}) \|^2 -\Upsilon\left(\frac{\lambda \| \boldsymbol{g}_{\mathbf{k}} \|}{2 \pi T_{ct}} \right) \nonumber \\
& \times \vert \boldsymbol{\Delta}_t(\mathbf{k}) \cdot \widehat{\boldsymbol{g}}_{\mathbf{k}} \vert^2 \bigg], \label{Eq_SC_OP05}
\end{align}
where $\Upsilon(x) \equiv \mathrm{Re}\left[\psi^{(0)}\left(\frac{1}{2} + i x \right) - \psi^{(0)}\left(\frac{1}{2} \right) \right]$, with $\psi^{(0)}(z)$ referring to the digamma function, and $\gamma = 0.57721...$ being the Euler constant. Notice that Eqs. \eqref{Eq_SC_OP04} and \eqref{Eq_SC_OP05} can be further simplified to
\begin{align}
\ln\left(\frac{T_{cs}}{T_{cs,0}}\right) & = - \int_{\mathcal{S}^2} \frac{d\Omega_{\mathbf{k}}}{4\pi} \Upsilon\left(\frac{\lambda \| \boldsymbol{g}_{\mathbf{k}} \|}{2 \pi T_{cs}} \right) \vert d_0(\mathbf{k}) \vert^2, \label{Eq_SC_OP06} \\
\ln\left(\frac{T_{ct}}{T_{ct,0}}\right) & = - \int_{\mathcal{S}^2} \frac{d\Omega_{\mathbf{k}}}{4\pi} \Upsilon\left(\frac{\lambda \| \boldsymbol{g}_{\mathbf{k}} \|}{2 \pi T_{ct}} \right) \vert \boldsymbol{d}(\mathbf{k}) \cdot \widehat{\boldsymbol{g}}_{\mathbf{k}} \vert^2, \label{Eq_SC_OP07}
\end{align}
where $\pi T_{cs, 0} = 2 e^{\gamma} \epsilon_c e^{-1/(N_F V_s)}$ and $\pi T_{ct, 0} = 2 e^{\gamma} \epsilon_c e^{-1/(N_F V_t)}$ define, respectively, the SC transition temperatures for the singlet and triplet channels in the absence of the AM interaction.

According to Eqs. \eqref{Eq_SC_OP06} and \eqref{Eq_SC_OP07}, the effect of the AM interaction on $T_{cs}$ and $T_{ct}$ depends on the function $\Upsilon(x)$, which is positive definite for $x \in [0, + \infty[$. For this reason, one expects that the increase of $\lambda$ results in the decay of $T_{cs}$ and $T_{ct}$, regardless of the angular momentum $\ell$ of the pairing gap function. We verified that this is true for the spin-singlet and spin-triplet gap functions with $s$-wave ($\ell = 0)$, $p$-wave ($\ell = 1)$, and $d$-wave ($\ell = 2)$ symmetries. This is illustrated in Fig. \ref{Tc_Altermag}, which shows the numerical solution of Eqs. \eqref{Eq_SC_OP06} and \eqref{Eq_SC_OP07} for spin-singlet ($\ell = 0, 2$) and spin-triplet ($\ell = 1$) gap functions belonging to the $D_{4h}$ point group with the $B^{-}_{1g}/D_{4h}$ form factor \footnote{The AM form factors $\boldsymbol{g}_{\mathbf{k}}$ used in the numerical calculations presented in this work are normalized at the Fermi surface.}. Notice that $s$-wave and $d$-wave pairing states are completely suppressed ($T_{cs} \rightarrow 0)$ as the AM interaction $\lambda$ becomes larger than a critical value. This behavior can be attributed to Anderson's theorem, since the AM interaction $\lambda$ breaks time-reversal symmetry \cite{Anderson-JPCS(1959)}. Moreover, although AM also tends to suppress $T_{ct}$ associated with $p$-wave pairing, this suppression tends to be milder. In fact, the decay of $T_{ct}$ as a function of $\lambda$ exhibits power-law behavior as $\lambda \gg T_{ct, 0}$.

There is also an infinite set of spin-triplet pairing states in which $T_{ct}$ is not suppressed at all by AM, but instead results in $T_{ct} = T_{ct,0}$ for any $\lambda$. In fact, according to Eq. \eqref{Eq_SC_OP05} such states must possess gap functions satisfying $\boldsymbol{d}(\mathbf{k}) \perp \boldsymbol{g}_{\mathbf{k}}$. Hence, they can be parametrized by $\boldsymbol{d}(\mathbf{k}) = \boldsymbol{u}(\mathbf{k}) \times \boldsymbol{g}_{\mathbf{k}}$, with $\boldsymbol{u}(\mathbf{k})$ being an odd-parity vector function in order to be consistent with Fermi-Dirac statistics. If we take the $p$-wave gap function $\boldsymbol{u}_{A_{1u}}(\mathbf{k}) = k_x \hat{\boldsymbol{x}} + k_y \hat{\boldsymbol{y}} + k_z \hat{\boldsymbol{z}}$ ($A_{1u}$ irrep of $D_{4h}$) and use the $B^{-}_{1g}/D_{4h}$ AM form factor, we obtain
\begin{equation}\label{fwave_GapFunc}
\boldsymbol{d}_{B_{2u}} = k_x (k^2_y - k^2_z) \hat{\boldsymbol{x}} + k_y (k^2_z - k^2_x) \hat{\boldsymbol{y}} + k_z (k^2_x - k^2_y) \hat{\boldsymbol{z}},
\end{equation}
which transforms as the $B_{2u}$ irrep of $D_{4h}$ and describes $f$-wave ($\ell = 3$) pairing. Employing other $p$-wave $\boldsymbol{u}(\mathbf{k})$ with the $B^{-}_{1g}/D_{4h}$ AM form factor will still result in $f$-wave gap functions. However, this is not generally valid for all AM states under the effect of SOC, because in some cases the components of their form factors do not have necessarily $d$-wave symmetry. The only feature which remains the same for the triplet states with $T_{ct} = T_{ct, 0}$ is the fact that their gap functions $\boldsymbol{d}(\mathbf{k}) = \boldsymbol{u}(\mathbf{k}) \times \boldsymbol{g}_{\mathbf{k}}$ will share the same nodal lines of the normal AM phase. Consequently, this constitutes a true fingerprint of the role played by altermagnetism in the promotion of such SC states.

\emph{Bogoliubov-Fermi surfaces and Weyl points.} Now, we investigate the effect of the AM interaction on the band structure of $\widehat{\mathcal{H}}_{\mathrm{BdG}}(\mathbf{k})$ [see Eq. \eqref{Eq_SC_Ham}]. First, notice that $\widehat{\mathcal{H}}_{\mathrm{BdG}}(\mathbf{k})$ has naturally charge-conjugation  symmetry $C$, since $\widehat{U}_C \widehat{\mathcal{H}}^{T}_{\mathrm{BdG}}(- \mathbf{k}) \widehat{U}^{\dagger}_C = - \widehat{\mathcal{H}}_{\mathrm{BdG}}(\mathbf{k})$, where $\widehat{U}_C = \hat{\tau}_y \otimes \hat{\sigma}_y$ and $\hat{\tau}_y$ is the second Pauli matrix in Nambu space. Consequently, its eigenvalues are given by $E_{\nu, \pm}(\mathbf{k}) = \pm E_{\nu}(\mathbf{k})$, where
\begin{equation}\label{Eq_SC_Disp}
\hspace{-0.3cm} E_{\nu}(\mathbf{k}) = 
\begin{cases}
\vert \sqrt{\xi^2_{\mathbf{k}} + \vert \Delta_s(\mathbf{k}) \vert^2} + \nu \lambda \| \boldsymbol{g}_{\mathbf{k}} \| \vert, \\
\sqrt{\xi^2_{\mathbf{k}} + \| \boldsymbol{\Delta}_t(\mathbf{k}) \|^2 + (\lambda \| \boldsymbol{g}_{\mathbf{k}} \|)^2 + \nu \phi(\mathbf{k})},  
\end{cases}
\end{equation}
$\nu = \pm$, and $\phi(\mathbf{k}) \equiv 2 \lambda \sqrt{\xi^2_{\mathbf{k}} \| \boldsymbol{g}_{\mathbf{k}} \|^2 + | \boldsymbol{\Delta}_t(\mathbf{k}) \cdot \boldsymbol{g}_{\mathbf{k}}|^2}$. The first (second) line of Eq. \eqref{Eq_SC_Disp} refers to the SC dispersions in a spin-singlet (spin-triplet) state. Interestingly, Eq. \eqref{Eq_SC_Disp} shows that in the presence of the AM interaction the nodal states of the spin-singlet SC form a Bogoliubov-Fermi surface \cite{Agterberg-PRL(2017),Brydon-PRB(2018)}. In this phase, the BdG Hamiltonian also has parity symmetry $P$, which when combined with $C$ results in $\widehat{U}_{CP} \widehat{\mathcal{H}}^T_{\mathrm{BdG}}(\mathbf{k}) \widehat{U}^{\dagger}_{CP} = - \widehat{\mathcal{H}}_{\mathrm{BdG}}(\mathbf{k})$, with $\widehat{U}_{CP} = \widehat{U}_C \widehat{U}^{*}_P = \hat{\tau}_y \otimes \hat{\sigma}_y$. Because $(CP)^2 \equiv (\widehat{U}_{CP} \mathcal{K})^2 = + \mathbb{1}$ and the fact that the system breaks time-reversal symmetry, the Bogoliubov-Fermi surface has a $\mathbb{Z}_2$ topological charge, which makes it stable against $CP$-preserving perturbations \cite{Kobayashi-PRB(2014),Zhao-PRL(2016)}. Indeed, following the same reasoning as in Refs. \cite{Agterberg-PRL(2017),Brydon-PRB(2018)}, one can show that there exists a unitary transformation $\widehat{\Omega}$ such that the Pfaffian $P(\mathbf{k}) \equiv \mathrm{Pf}[\widehat{\Omega} \widehat{\mathcal{H}}_{\mathrm{BdG}}(\mathbf{k}) \widehat{\Omega}^{\dagger}]$ becomes $P(\mathbf{k}) = \xi^2_{\mathbf{k}} + \vert \Delta_s(\mathbf{k}) \vert^2 - (\lambda \| \boldsymbol{g}_{\mathbf{k}} \|)^2$ (see SM \cite{Suppl_Mat}). Since $P(\mathbf{k})$ changes sign at the Bogoliubov-Fermi surface, the $\mathbb{Z}_2$ topological invariant assumes the form $(-1)^{l} = \mathrm{sgn}[P(\mathbf{k}_{+}) P(\mathbf{k}_{-})]$, where $\mathbf{k}_{+}$ and $\mathbf{k}_{-}$ refer, respectively, to momenta outside and inside of this surface. Remarkably, this $\mathbb{Z}_2$ invariant also characterizes the Fermi surface of the metallic AM phase, since in this case $P(\mathbf{k})$ also features a sign change as $\mathbf{k}$ crosses the Fermi momentum.

The spin-triplet SC phase does not allow extended Bogoliubov-Fermi surfaces because it breaks the parity symmetry. In fact, the analysis of the dispersions in Eq. \eqref{Eq_SC_Disp} shows that the spin-triplet band structure exhibits either gapped or nodal-point excitations. The former appears when the gap function $\boldsymbol{d}(\mathbf{k})$ does not possess the mirror symmetry of the AM phase, which ensures the existence of the nodal lines. In contrast, if $\boldsymbol{d}(\mathbf{k})$ shares at least one of the mirror plane of the AM spectrum, the nodal structures in the spin-triplet SC phase will exhibit Weyl-like excitations. This is particularly clear for the spin-triplet states $\boldsymbol{d}(\mathbf{k}) = \boldsymbol{u}(\mathbf{k}) \times \boldsymbol{g}_{\mathbf{k}}$, which feature the dispersions $E_{\pm}(\mathbf{k}) = \sqrt{(\vert \xi_{\mathbf{k}}\vert \pm \lambda \|\boldsymbol{g}_{\mathbf{k}} \|)^2 + \| \Delta_t \boldsymbol{d}(\mathbf{k}) \|^2}$. Therefore, the gapless excitations for this phase are comprised of Weyl-pinching points along the nodal lines $\boldsymbol{g}_{\mathbf{k}} = \boldsymbol{0}$ and $\boldsymbol{u}(\mathbf{k}) \times \boldsymbol{g}_{\mathbf{k}} = \boldsymbol{0}$. In addition, these nodal lines are described by $\pm \pi$ Berry phases (see SM \cite{Suppl_Mat}), which thus make the spin-triplet SC phase topologically non-trivial.

\begin{figure}[t]
\centering
\centering \includegraphics[width=1.0\linewidth,valign=t]{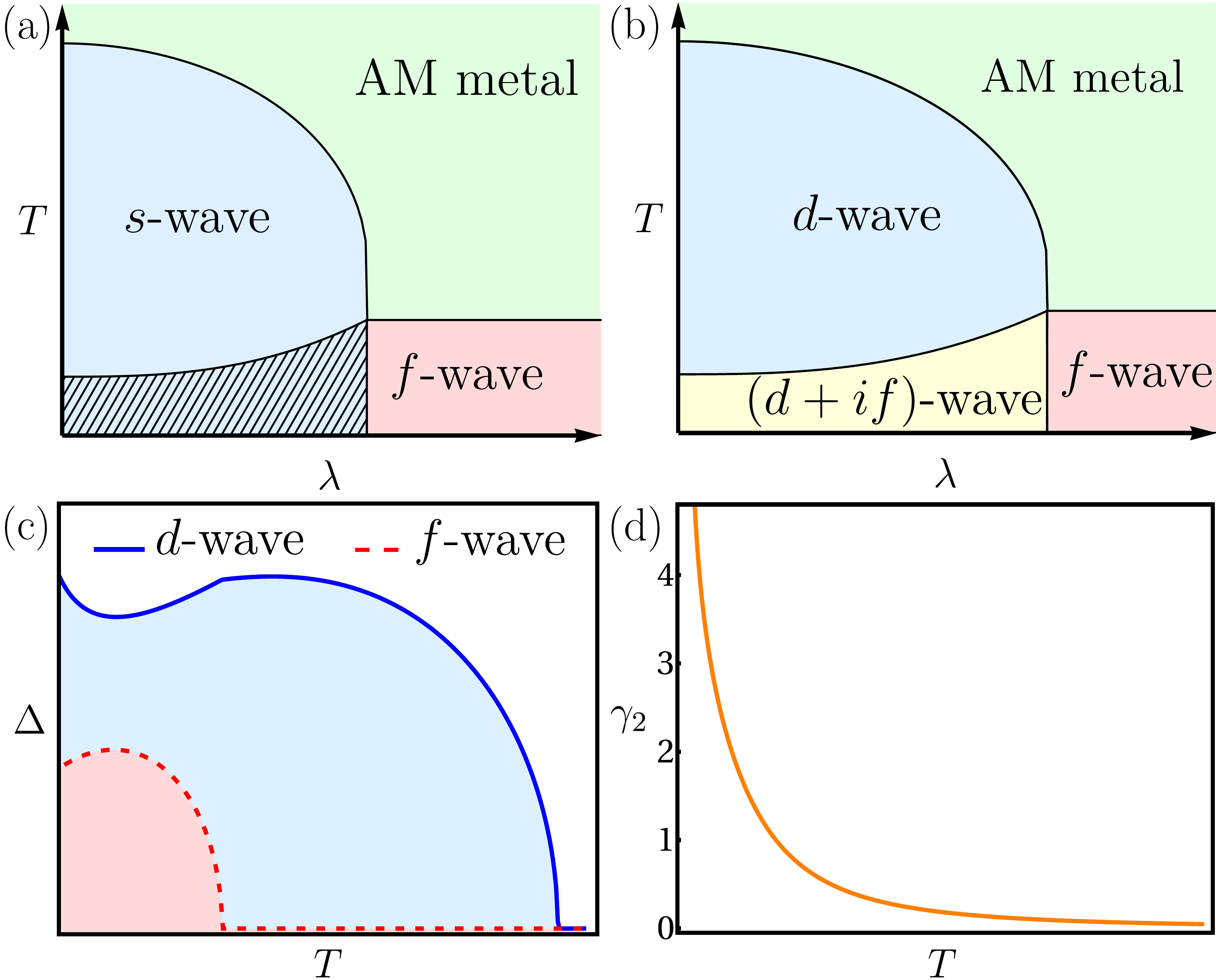}
\caption{(a), (b) Schematic phase diagrams of temperature ($T$) versus AM interaction ($\lambda$) illustrating two scenarios for the emergence of singlet and triplet pairings in a $B^{-}_{1g}/D_{4h}$ AM metal. In the hatched area in (a), the minimization of the GL free energy yields divergent SC order parameters, making it much harder to establish the coexistence between the $s$- and $f$-wave SC phases. (b) In contrast, the $d_{x^2 - y^2}$- and $f$-wave SC phases coexist within a finite range of $\lambda$. (c) This is signaled by the non-zero values assumed by the SC amplitudes $\Delta_s$ and $\Delta_t$ as $T$ decreases. (d) Since the GL coefficient $\gamma_2$ is always positive, this coexisting phase breaks the time-reversal invariance, resulting in a chiral $d + i f$ state.}\label{AM_SC_PhaseDiag}
\end{figure}

\emph{Ginzburg-Landau analysis.} As we have seen earlier, AM phases resulting from SOC are compatible with both spin-singlet and spin-triplet superconductivity emerging at low temperatures. In that situation, the conventional wisdom is that the first developing SC order dominates the subleading one by generating negative feedback \cite{Fernandes-PRB(2012),Hinojosa-PRL(2014)}. However, if the repulsion between those competing orders is not too large, there is also a possibility of their coexistence at low enough temperatures. To address this issue, we now investigate the properties of the Ginzburg-Landau (GL) free energy $\mathcal{F}[\Delta_s, \Delta_t]$ for this coexisting SC phase \cite{Altland-CUP(2010)}. As explained in the SM \cite{Suppl_Mat}, after integrating out the fermionic degrees of freedom of the SC action related to the BdG Hamiltonian, we obtain that the expansion of $\mathcal{F}[\Delta_s, \Delta_t]$ up to fourth order in the order parameters $\Delta_s$ and $\Delta_t$ yields:
\begin{align}\label{Eq_FreeEner}
\mathcal{F} & = \mathcal{F}_0 + \alpha_1 \vert \Delta_s \vert^2 + \alpha_2 \vert \Delta_t \vert^2 + \beta_1 \vert \Delta_s \vert^4 +\beta_2 \vert \Delta_t \vert^4 \nonumber\\
&+ 2 \gamma_1 \vert \Delta_s \vert^2 \vert \Delta_t \vert^2 + \gamma_2 [ \Delta^2_s (\Delta^{*}_t)^2 + (\Delta^{*}_s)^2 \Delta^2_t],
\end{align}
where $\mathcal{F}_0$ represents the normal state contribution and the coefficients multiplying $\Delta_s$ and $\Delta_t$ depend on the temperature, the AM interaction $\lambda$, and the form factor $\boldsymbol{g}_{\mathbf{k}}$. Naturally, the minimization of $\mathcal{F}[\Delta_s, \Delta_t]$ also depends on the phase difference $\theta$ between $\Delta_s = \vert \Delta_s \vert e^{i(\Phi + \theta/2)}$ and $\Delta_t = \vert \Delta_t \vert e^{i(\Phi - \theta/2)}$, which can be determined by the sign of the GL coefficient $\gamma_2$. For the situation in which a spin-singlet SC state coexists with a spin-triplet state defined by the gap function $\boldsymbol{d}(\mathbf{k}) = \boldsymbol{u}(\mathbf{k}) \times \boldsymbol{g}_{\mathbf{k}}$, this coefficient becomes: 
\begin{align}
\gamma_2 & = \frac{N_F}{4} \int_{\mathcal{S}^2} \frac{d\Omega_{\mathbf{k}}}{4 \pi} \frac{d^2_0(\mathbf{k}) \boldsymbol{d}^2(\mathbf{k})}{(\lambda \| \boldsymbol{g}_{\mathbf{k}} \|)^2} \Upsilon \left(\frac{\lambda \| \boldsymbol{g}_{\mathbf{k}} \|}{2 \pi T} \right).
\end{align}
Since the integrand in the above equation depends on the positive function $\Upsilon(x)$, $\gamma_2$ turns out to be always positive regardless of the parent AM phase. Consequently, the minimization of the free energy $\mathcal{F}[\Delta_s, \Delta_t]$ yields the solution $\theta = \pm \pi/2$. Since these phase values represent different states, the SC phase breaks spontaneously the time-reversal symmetry \cite{Wu-PRL(2009),Wu-PRB(2010),Chubukov-PRB(2013),Chubukov-PRB(2014),Hinojosa-PRL(2014),Wang-PRL(2017)}, which was already broken in the AM state. Nonetheless, the validity of this reasoning is based on finding $\vert \Delta_s \vert > 0$ and $\vert \Delta_t \vert > 0$ below a certain critical temperature. In fact, the necessary and sufficient conditions for the emergence of such a coexisting SC phase imply that the GL coefficients must obey the inequalities $\beta_1 \beta_2 > (\gamma_1 - \gamma_2)^2$, $\alpha_2 (\gamma_1 - \gamma_2) - \alpha_1 \beta_2 > 0$, and $\alpha_1 (\gamma_1 - \gamma_2) - \alpha_2 \beta_1 > 0$, as can be demonstrated from the minimization of $\mathcal{F}[\Delta_s, \Delta_t]$ \cite{Hinojosa-PRL(2014)}.

To address this point, we investigate the coexistence of the spin-triplet state defined by the $f$-wave gap function $\boldsymbol{d}_{B_{2u}}(\mathbf{k})$ with spin-singlet states possessing either $s$-wave or $d_{x^2 - y^2}$-wave symmetries inside an AM phase with $B^{-}_{1g}/D_{4h}$ order parameter \footnote{The $d_{xy}$-wave SC state was not taken into account because it has a lower critical temperature compared to the $s$-wave and $d_{x^2 - y^2}$-wave SC states, when the AM interaction is switched on.}. Since the GL coefficients are functions of $T$ and $\lambda$, these two parameters become the only variables controlling the minimization of $\mathcal{F}[\Delta_s, \Delta_t]$. We found that the minimization of the GL free energy in Eq. \eqref{Eq_FreeEner} does not support a coexisting state between the $s$-wave and the $f$-wave SC phases, because the SC order parameters go to infinity at low temperatures, signaling the breakdown of the GL expansion \footnote{However, we cannot rule out the possibility that this picture could change with the inclusion of higher-order terms in the expansion of the GL free energy. The latter study is nevertheless beyond the scope of the present work.}. On the other hand, we do obtain coexistence between the $d_{x^2 - y^2}$-wave and the $f$-wave SC states at low temperatures within a finite range of AM interactions (see Fig. \ref{AM_SC_PhaseDiag}). As the SC state must break time-reversal invariance in this phase, this leads to a chiral $d + if$ state. We should point out that this last result relies on the fact that the AM phase from which superconductivity emerges is described by an order parameter transforming as the $B^{-}_{1g}$ irrep of the $D_{4h}$ point group. Had we started from an AM phase with an order parameter defined in terms of a higher angular-moment form factor, the spin-triplet component of the resulting chiral SC state would have a larger symmetry than the $f$-wave SC state. For instance, this would occur for an AM phase in a cubic system in which the order parameter transforms as the $A^{-}_{1g}$ irrep of $O_h$ \cite{Fernandes-PRB(2024)}. The components of the form factor for this system possess $g$-wave symmetry. As a result, the simplest spin-triplet state allowed by the AM interaction [i.e., with gap function $\boldsymbol{d}(\mathbf{k}) = \boldsymbol{u}(\mathbf{k}) \times \boldsymbol{g}_{\mathbf{k}}$] that can potentially coexist with a $d$-wave spin-singlet state occurs in the $h$-wave channel.

\emph{Summary and outlook.} We have studied here some of the possibilities for pairing states in 3D models for SOC altermagnets \cite{Fernandes-PRB(2024)}. For a generic pairing interaction that favors both spin-singlet and spin-triplet superconductivity, we found that altermagnetism is beneficial to the latter over the former. In fact, the gap function of the favored triplet SC state inherits the same nodal-line structure and the topological properties of the AM order parameter. As a result, this makes $f$-wave triplet superconductivity as the most likely partner to a $d$-wave AM phase. In addition, the combination of AM and singlet superconductivity favors the appearance of topological $\mathbb{Z}_2$ Bogoliubov-Fermi surfaces, which cannot be gapped out by $CP$-preserving perturbations. Through a Ginzburg-Landau analysis, we have also established in AM metals with $B^{-}_{1g}/D_{4h}$ symmetry the emergence of a coexisting phase between $f$-wave and $d_{x^2 - y^2}$-wave superconductivity, which breaks time-reversal symmetry spontaneously. We argue that these features are ubiquitous and should appear in other families of altermagnets in which the SOC is relevant.

\emph{Acknowledgments.} We would like to thank interesting conversations with Rafael M. Fernandes and Rodrigo G. Pereira. We acknowledge funding from the Brazilian agency Conselho Nacional de Desenvolvimento Cient\'{\i}fico e Tecnol\'ogico (CNPq) under grant No. 404274/2023-4. H.F. also acknowledges funding from CNPq under grant No. 311428/2021-5.


\begin{thebibliography}{53}%
\makeatletter
\providecommand \@ifxundefined [1]{%
 \@ifx{#1\undefined}
}%
\providecommand \@ifnum [1]{%
 \ifnum #1\expandafter \@firstoftwo
 \else \expandafter \@secondoftwo
 \fi
}%
\providecommand \@ifx [1]{%
 \ifx #1\expandafter \@firstoftwo
 \else \expandafter \@secondoftwo
 \fi
}%
\providecommand \natexlab [1]{#1}%
\providecommand \enquote  [1]{``#1''}%
\providecommand \bibnamefont  [1]{#1}%
\providecommand \bibfnamefont [1]{#1}%
\providecommand \citenamefont [1]{#1}%
\providecommand \href@noop [0]{\@secondoftwo}%
\providecommand \href [0]{\begingroup \@sanitize@url \@href}%
\providecommand \@href[1]{\@@startlink{#1}\@@href}%
\providecommand \@@href[1]{\endgroup#1\@@endlink}%
\providecommand \@sanitize@url [0]{\catcode `\\12\catcode `\$12\catcode
  `\&12\catcode `\#12\catcode `\^12\catcode `\_12\catcode `\%12\relax}%
\providecommand \@@startlink[1]{}%
\providecommand \@@endlink[0]{}%
\providecommand \url  [0]{\begingroup\@sanitize@url \@url }%
\providecommand \@url [1]{\endgroup\@href {#1}{\urlprefix }}%
\providecommand \urlprefix  [0]{URL }%
\providecommand \Eprint [0]{\href }%
\providecommand \doibase [0]{http://dx.doi.org/}%
\providecommand \selectlanguage [0]{\@gobble}%
\providecommand \bibinfo  [0]{\@secondoftwo}%
\providecommand \bibfield  [0]{\@secondoftwo}%
\providecommand \translation [1]{[#1]}%
\providecommand \BibitemOpen [0]{}%
\providecommand \bibitemStop [0]{}%
\providecommand \bibitemNoStop [0]{.\EOS\space}%
\providecommand \EOS [0]{\spacefactor3000\relax}%
\providecommand \BibitemShut  [1]{\csname bibitem#1\endcsname}%
\let\auto@bib@innerbib\@empty
\bibitem [{\citenamefont {\ifmmode~\check{S}\else \v{S}\fi{}mejkal}\ \emph
  {et~al.}(2022{\natexlab{a}})\citenamefont {\ifmmode~\check{S}\else
  \v{S}\fi{}mejkal}, \citenamefont {Sinova},\ and\ \citenamefont
  {Jungwirth}}]{Smejkal2022_1}%
  \BibitemOpen
  \bibfield  {author} {\bibinfo {author} {\bibfnamefont {L.}~\bibnamefont
  {\ifmmode~\check{S}\else \v{S}\fi{}mejkal}}, \bibinfo {author} {\bibfnamefont
  {J.}~\bibnamefont {Sinova}}, \ and\ \bibinfo {author} {\bibfnamefont
  {T.}~\bibnamefont {Jungwirth}},\ }\bibfield  {title} {\bibinfo {title}
  {{Beyond Conventional Ferromagnetism and Antiferromagnetism: A Phase with
  Nonrelativistic Spin and Crystal Rotation Symmetry}},\ }\href {\doibase
  10.1103/PhysRevX.12.031042} {\bibfield  {journal} {\bibinfo  {journal} {Phys.
  Rev. X}\ }\textbf {\bibinfo {volume} {12}},\ \bibinfo {pages} {031042}
  (\bibinfo {year} {2022}{\natexlab{a}})}\BibitemShut {NoStop}%
\bibitem [{\citenamefont {\ifmmode~\check{S}\else \v{S}\fi{}mejkal}\ \emph
  {et~al.}(2022{\natexlab{b}})\citenamefont {\ifmmode~\check{S}\else
  \v{S}\fi{}mejkal}, \citenamefont {Sinova},\ and\ \citenamefont
  {Jungwirth}}]{Smejkal2022_2}%
  \BibitemOpen
  \bibfield  {author} {\bibinfo {author} {\bibfnamefont {L.}~\bibnamefont
  {\ifmmode~\check{S}\else \v{S}\fi{}mejkal}}, \bibinfo {author} {\bibfnamefont
  {J.}~\bibnamefont {Sinova}}, \ and\ \bibinfo {author} {\bibfnamefont
  {T.}~\bibnamefont {Jungwirth}},\ }\bibfield  {title} {\bibinfo {title}
  {{Emerging Research Landscape of Altermagnetism}},\ }\href {\doibase
  10.1103/PhysRevX.12.040501} {\bibfield  {journal} {\bibinfo  {journal} {Phys.
  Rev. X}\ }\textbf {\bibinfo {volume} {12}},\ \bibinfo {pages} {040501}
  (\bibinfo {year} {2022}{\natexlab{b}})}\BibitemShut {NoStop}%
\bibitem [{\citenamefont {Jungwirth}\ \emph {et~al.}(2024)\citenamefont
  {Jungwirth}, \citenamefont {Fernandes}, \citenamefont {Sinova},\ and\
  \citenamefont {\v{S}mejkal}}]{Jungwirth-arXiv(2024)}%
  \BibitemOpen
  \bibfield  {author} {\bibinfo {author} {\bibfnamefont {T.}~\bibnamefont
  {Jungwirth}}, \bibinfo {author} {\bibfnamefont {R.~M.}\ \bibnamefont
  {Fernandes}}, \bibinfo {author} {\bibfnamefont {J.}~\bibnamefont {Sinova}}, \
  and\ \bibinfo {author} {\bibfnamefont {L.}~\bibnamefont {\v{S}mejkal}},\
  }\bibfield  {title} {\bibinfo {title} {{Altermagnets and beyond: Nodal
  magnetically-ordered phases}},\ }\href {https://arxiv.org/abs/2409.10034}
  {\bibfield  {journal} {\bibinfo  {journal} {arXiv:2409.10034}\ } (\bibinfo
  {year} {2024})}\BibitemShut {NoStop}%
\bibitem [{\citenamefont {Fedchenko}\ \emph {et~al.}(2024)\citenamefont
  {Fedchenko}, \citenamefont {Min\'ar}, \citenamefont {Akashdeep},
  \citenamefont {D'Souza}, \citenamefont {Vasilyev}, \citenamefont {Tkach},
  \citenamefont {Odenbreit}, \citenamefont {Nguyen}, \citenamefont
  {Kutnyakhov}, \citenamefont {Wind}, \citenamefont {Wenthaus}, \citenamefont
  {Scholz}, \citenamefont {Rossnagel}, \citenamefont {Hoesch}, \citenamefont
  {Aeschlimann}, \citenamefont {Stadtm\"uller}, \citenamefont {Kl\"aui},
  \citenamefont {Sch\"onhense}, \citenamefont {Jungwirth}, \citenamefont
  {Hellenes}, \citenamefont {Jakob}, \citenamefont {\v{S}mejkal}, \citenamefont
  {Sinova},\ and\ \citenamefont {Elmers}}]{Fedchenko_2024}%
  \BibitemOpen
  \bibfield  {author} {\bibinfo {author} {\bibfnamefont {O.}~\bibnamefont
  {Fedchenko}}, \bibinfo {author} {\bibfnamefont {J.}~\bibnamefont {Min\'ar}},
  \bibinfo {author} {\bibfnamefont {A.}~\bibnamefont {Akashdeep}}, \bibinfo
  {author} {\bibfnamefont {S.~W.}\ \bibnamefont {D'Souza}}, \bibinfo {author}
  {\bibfnamefont {D.}~\bibnamefont {Vasilyev}}, \bibinfo {author}
  {\bibfnamefont {O.}~\bibnamefont {Tkach}}, \bibinfo {author} {\bibfnamefont
  {L.}~\bibnamefont {Odenbreit}}, \bibinfo {author} {\bibfnamefont
  {Q.}~\bibnamefont {Nguyen}}, \bibinfo {author} {\bibfnamefont
  {D.}~\bibnamefont {Kutnyakhov}}, \bibinfo {author} {\bibfnamefont
  {N.}~\bibnamefont {Wind}}, \bibinfo {author} {\bibfnamefont {L.}~\bibnamefont
  {Wenthaus}}, \bibinfo {author} {\bibfnamefont {M.}~\bibnamefont {Scholz}},
  \bibinfo {author} {\bibfnamefont {K.}~\bibnamefont {Rossnagel}}, \bibinfo
  {author} {\bibfnamefont {M.}~\bibnamefont {Hoesch}}, \bibinfo {author}
  {\bibfnamefont {M.}~\bibnamefont {Aeschlimann}}, \bibinfo {author}
  {\bibfnamefont {B.}~\bibnamefont {Stadtm\"uller}}, \bibinfo {author}
  {\bibfnamefont {M.}~\bibnamefont {Kl\"aui}}, \bibinfo {author} {\bibfnamefont
  {G.}~\bibnamefont {Sch\"onhense}}, \bibinfo {author} {\bibfnamefont
  {T.}~\bibnamefont {Jungwirth}}, \bibinfo {author} {\bibfnamefont {A.~B.}\
  \bibnamefont {Hellenes}}, \bibinfo {author} {\bibfnamefont {G.}~\bibnamefont
  {Jakob}}, \bibinfo {author} {\bibfnamefont {L.}~\bibnamefont {\v{S}mejkal}},
  \bibinfo {author} {\bibfnamefont {J.}~\bibnamefont {Sinova}}, \ and\ \bibinfo
  {author} {\bibfnamefont {H.-J.}\ \bibnamefont {Elmers}},\ }\bibfield  {title}
  {\bibinfo {title} {{Observation of time-reversal symmetry breaking in the
  band structure of altermagnetic $\mathrm{Ru}\mathrm{O}_2$}},\ }\href
  {\doibase 10.1126/sciadv.adj4883} {\bibfield  {journal} {\bibinfo  {journal}
  {Sci. Adv.}\ }\textbf {\bibinfo {volume} {10}},\ \bibinfo {pages} {eadj4883}
  (\bibinfo {year} {2024})}\BibitemShut {NoStop}%
\bibitem [{\citenamefont {Lee}\ \emph {et~al.}(2024)\citenamefont {Lee},
  \citenamefont {Lee}, \citenamefont {Jung}, \citenamefont {Jung},
  \citenamefont {Kim}, \citenamefont {Lee}, \citenamefont {Seok}, \citenamefont
  {Kim}, \citenamefont {Park}, \citenamefont {\ifmmode~\check{S}\else
  \v{S}\fi{}mejkal}, \citenamefont {Kang},\ and\ \citenamefont
  {Kim}}]{Libor_MnTe}%
  \BibitemOpen
  \bibfield  {author} {\bibinfo {author} {\bibfnamefont {S.}~\bibnamefont
  {Lee}}, \bibinfo {author} {\bibfnamefont {S.}~\bibnamefont {Lee}}, \bibinfo
  {author} {\bibfnamefont {S.}~\bibnamefont {Jung}}, \bibinfo {author}
  {\bibfnamefont {J.}~\bibnamefont {Jung}}, \bibinfo {author} {\bibfnamefont
  {D.}~\bibnamefont {Kim}}, \bibinfo {author} {\bibfnamefont {Y.}~\bibnamefont
  {Lee}}, \bibinfo {author} {\bibfnamefont {B.}~\bibnamefont {Seok}}, \bibinfo
  {author} {\bibfnamefont {J.}~\bibnamefont {Kim}}, \bibinfo {author}
  {\bibfnamefont {B.~G.}\ \bibnamefont {Park}}, \bibinfo {author}
  {\bibfnamefont {L.}~\bibnamefont {\ifmmode~\check{S}\else \v{S}\fi{}mejkal}},
  \bibinfo {author} {\bibfnamefont {C.-J.}\ \bibnamefont {Kang}}, \ and\
  \bibinfo {author} {\bibfnamefont {C.}~\bibnamefont {Kim}},\ }\bibfield
  {title} {\bibinfo {title} {{Broken Kramers Degeneracy in Altermagnetic
  MnTe}},\ }\href {\doibase 10.1103/PhysRevLett.132.036702} {\bibfield
  {journal} {\bibinfo  {journal} {Phys. Rev. Lett.}\ }\textbf {\bibinfo
  {volume} {132}},\ \bibinfo {pages} {036702} (\bibinfo {year}
  {2024})}\BibitemShut {NoStop}%
\bibitem [{\citenamefont {\v{S}mejkal}\ \emph {et~al.}(2020)\citenamefont
  {\v{S}mejkal}, \citenamefont {Gonz\'alez-Hern\'andez}, \citenamefont
  {Jungwirth},\ and\ \citenamefont {Sinova}}]{Smejkal2022_3}%
  \BibitemOpen
  \bibfield  {author} {\bibinfo {author} {\bibfnamefont {L.}~\bibnamefont
  {\v{S}mejkal}}, \bibinfo {author} {\bibfnamefont {R.}~\bibnamefont
  {Gonz\'alez-Hern\'andez}}, \bibinfo {author} {\bibfnamefont {T.}~\bibnamefont
  {Jungwirth}}, \ and\ \bibinfo {author} {\bibfnamefont {J.}~\bibnamefont
  {Sinova}},\ }\bibfield  {title} {\bibinfo {title} {Crystal time-reversal
  symmetry breaking and spontaneous hall effect in collinear
  antiferromagnets},\ }\href {\doibase 10.1126/sciadv.aaz8809} {\bibfield
  {journal} {\bibinfo  {journal} {Sci. Adv.}\ }\textbf {\bibinfo {volume}
  {6}},\ \bibinfo {pages} {eaaz8809} (\bibinfo {year} {2020})}\BibitemShut
  {NoStop}%
\bibitem [{\citenamefont {Mazin}\ \emph {et~al.}(2021)\citenamefont {Mazin},
  \citenamefont {Koepernik}, \citenamefont {Johannes}, \citenamefont
  {Gonz\'alez-Hern\'andez},\ and\ \citenamefont {\v{S}mejkal}}]{Mazin_PNAS}%
  \BibitemOpen
  \bibfield  {author} {\bibinfo {author} {\bibfnamefont {I.~I.}\ \bibnamefont
  {Mazin}}, \bibinfo {author} {\bibfnamefont {K.}~\bibnamefont {Koepernik}},
  \bibinfo {author} {\bibfnamefont {M.~D.}\ \bibnamefont {Johannes}}, \bibinfo
  {author} {\bibfnamefont {R.~R.}\ \bibnamefont {Gonz\'alez-Hern\'andez}}, \
  and\ \bibinfo {author} {\bibfnamefont {L.}~\bibnamefont {\v{S}mejkal}},\
  }\bibfield  {title} {\bibinfo {title} {{Prediction of unconventional
  magnetism in doped $\mathrm{Fe}\mathrm{Sb}_2$}},\ }\href {\doibase
  10.1073/pnas.2108924118} {\bibfield  {journal} {\bibinfo  {journal} {Proc.
  Natl. Acad. Sci. U.S.A.}\ }\textbf {\bibinfo {volume} {118}},\ \bibinfo
  {pages} {e2108924118} (\bibinfo {year} {2021})}\BibitemShut {NoStop}%
\bibitem [{\citenamefont {Mazin}(2022)}]{Mazin2022notes}%
  \BibitemOpen
  \bibfield  {author} {\bibinfo {author} {\bibfnamefont {I.~I.}\ \bibnamefont
  {Mazin}},\ }\bibfield  {title} {\bibinfo {title} {{Notes on altermagnetism
  and superconductivity}},\ }\href {https://arxiv.org/abs/2203.05000}
  {\bibfield  {journal} {\bibinfo  {journal} {arXiv:2203.05000}\ } (\bibinfo
  {year} {2022})}\BibitemShut {NoStop}%
\bibitem [{\citenamefont {Sigrist}\ and\ \citenamefont
  {Ueda}(1991)}]{Sigrist-RMP(1991)}%
  \BibitemOpen
  \bibfield  {author} {\bibinfo {author} {\bibfnamefont {M.}~\bibnamefont
  {Sigrist}}\ and\ \bibinfo {author} {\bibfnamefont {K.}~\bibnamefont {Ueda}},\
  }\bibfield  {title} {\bibinfo {title} {{Phenomenological theory of
  unconventional superconductivity}},\ }\href {\doibase
  10.1103/RevModPhys.63.239} {\bibfield  {journal} {\bibinfo  {journal} {Rev.
  Mod. Phys.}\ }\textbf {\bibinfo {volume} {63}},\ \bibinfo {pages} {239}
  (\bibinfo {year} {1991})}\BibitemShut {NoStop}%
\bibitem [{\citenamefont {Scalapino}\ \emph {et~al.}(1986)\citenamefont
  {Scalapino}, \citenamefont {Loh},\ and\ \citenamefont
  {Hirsch}}]{Scalapino_1986}%
  \BibitemOpen
  \bibfield  {author} {\bibinfo {author} {\bibfnamefont {D.~J.}\ \bibnamefont
  {Scalapino}}, \bibinfo {author} {\bibfnamefont {E.}~\bibnamefont {Loh}}, \
  and\ \bibinfo {author} {\bibfnamefont {J.~E.}\ \bibnamefont {Hirsch}},\
  }\bibfield  {title} {\bibinfo {title} {{$d$-wave pairing near a
  spin-density-wave instability}},\ }\href {\doibase 10.1103/PhysRevB.34.8190}
  {\bibfield  {journal} {\bibinfo  {journal} {Phys. Rev. B}\ }\textbf {\bibinfo
  {volume} {34}},\ \bibinfo {pages} {8190} (\bibinfo {year}
  {1986})}\BibitemShut {NoStop}%
\bibitem [{\citenamefont {B\'eal-Monod}\ \emph {et~al.}(1986)\citenamefont
  {B\'eal-Monod}, \citenamefont {Bourbonnais},\ and\ \citenamefont
  {Emery}}]{Monod_Emery}%
  \BibitemOpen
  \bibfield  {author} {\bibinfo {author} {\bibfnamefont {M.~T.}\ \bibnamefont
  {B\'eal-Monod}}, \bibinfo {author} {\bibfnamefont {C.}~\bibnamefont
  {Bourbonnais}}, \ and\ \bibinfo {author} {\bibfnamefont {V.~J.}\ \bibnamefont
  {Emery}},\ }\bibfield  {title} {\bibinfo {title} {{Possible superconductivity
  in nearly antiferromagnetic itinerant fermion systems}},\ }\href {\doibase
  10.1103/PhysRevB.34.7716} {\bibfield  {journal} {\bibinfo  {journal} {Phys.
  Rev. B}\ }\textbf {\bibinfo {volume} {34}},\ \bibinfo {pages} {7716}
  (\bibinfo {year} {1986})}\BibitemShut {NoStop}%
\bibitem [{\citenamefont {Miyake}\ \emph {et~al.}(1986)\citenamefont {Miyake},
  \citenamefont {Schmitt-Rink},\ and\ \citenamefont {Varma}}]{Miyake_Varma}%
  \BibitemOpen
  \bibfield  {author} {\bibinfo {author} {\bibfnamefont {K.}~\bibnamefont
  {Miyake}}, \bibinfo {author} {\bibfnamefont {S.}~\bibnamefont
  {Schmitt-Rink}}, \ and\ \bibinfo {author} {\bibfnamefont {C.~M.}\
  \bibnamefont {Varma}},\ }\bibfield  {title} {\bibinfo {title}
  {{Spin-fluctuation-mediated even-parity pairing in heavy-fermion
  superconductors}},\ }\href {\doibase 10.1103/PhysRevB.34.6554} {\bibfield
  {journal} {\bibinfo  {journal} {Phys. Rev. B}\ }\textbf {\bibinfo {volume}
  {34}},\ \bibinfo {pages} {6554} (\bibinfo {year} {1986})}\BibitemShut
  {NoStop}%
\bibitem [{\citenamefont {Fay}\ and\ \citenamefont
  {Layzer}(1968)}]{Fay_Layzer}%
  \BibitemOpen
  \bibfield  {author} {\bibinfo {author} {\bibfnamefont {D.}~\bibnamefont
  {Fay}}\ and\ \bibinfo {author} {\bibfnamefont {A.}~\bibnamefont {Layzer}},\
  }\bibfield  {title} {\bibinfo {title} {{Superfluidity of Low-Density Fermion
  Systems}},\ }\href {\doibase 10.1103/PhysRevLett.20.187} {\bibfield
  {journal} {\bibinfo  {journal} {Phys. Rev. Lett.}\ }\textbf {\bibinfo
  {volume} {20}},\ \bibinfo {pages} {187} (\bibinfo {year} {1968})}\BibitemShut
  {NoStop}%
\bibitem [{\citenamefont {Berk}\ and\ \citenamefont
  {Schrieffer}(1966)}]{Berk_Schrieffer}%
  \BibitemOpen
  \bibfield  {author} {\bibinfo {author} {\bibfnamefont {N.~F.}\ \bibnamefont
  {Berk}}\ and\ \bibinfo {author} {\bibfnamefont {J.~R.}\ \bibnamefont
  {Schrieffer}},\ }\bibfield  {title} {\bibinfo {title} {{Effect of
  Ferromagnetic Spin Correlations on Superconductivity}},\ }\href {\doibase
  10.1103/PhysRevLett.17.433} {\bibfield  {journal} {\bibinfo  {journal} {Phys.
  Rev. Lett.}\ }\textbf {\bibinfo {volume} {17}},\ \bibinfo {pages} {433}
  (\bibinfo {year} {1966})}\BibitemShut {NoStop}%
\bibitem [{\citenamefont {Schiff}\ \emph {et~al.}(2024)\citenamefont {Schiff},
  \citenamefont {Corticelli}, \citenamefont {Guerreiro}, \citenamefont
  {Romh\'anyi},\ and\ \citenamefont {McClarty}}]{Schiff-arXiv(2024)}%
  \BibitemOpen
  \bibfield  {author} {\bibinfo {author} {\bibfnamefont {H.}~\bibnamefont
  {Schiff}}, \bibinfo {author} {\bibfnamefont {A.}~\bibnamefont {Corticelli}},
  \bibinfo {author} {\bibfnamefont {A.}~\bibnamefont {Guerreiro}}, \bibinfo
  {author} {\bibfnamefont {J.}~\bibnamefont {Romh\'anyi}}, \ and\ \bibinfo
  {author} {\bibfnamefont {P.}~\bibnamefont {McClarty}},\ }\bibfield  {title}
  {\bibinfo {title} {{The Spin Point Groups and their Representations}},\
  }\href {https://arxiv.org/abs/2307.12784} {\bibfield  {journal} {\bibinfo
  {journal} {arXiv:2307.12784}\ } (\bibinfo {year} {2024})}\BibitemShut
  {NoStop}%
\bibitem [{\citenamefont {McClarty}\ and\ \citenamefont
  {Rau}(2024)}]{McClarty-PRL(2024)}%
  \BibitemOpen
  \bibfield  {author} {\bibinfo {author} {\bibfnamefont {P.~A.}\ \bibnamefont
  {McClarty}}\ and\ \bibinfo {author} {\bibfnamefont {J.~G.}\ \bibnamefont
  {Rau}},\ }\bibfield  {title} {\bibinfo {title} {Landau theory of
  altermagnetism},\ }\href {\doibase 10.1103/PhysRevLett.132.176702} {\bibfield
   {journal} {\bibinfo  {journal} {Phys. Rev. Lett.}\ }\textbf {\bibinfo
  {volume} {132}},\ \bibinfo {pages} {176702} (\bibinfo {year}
  {2024})}\BibitemShut {NoStop}%
\bibitem [{\citenamefont {Ouassou}\ \emph {et~al.}(2023)\citenamefont
  {Ouassou}, \citenamefont {Brataas},\ and\ \citenamefont
  {Linder}}]{Ouassou-PRL(2023)}%
  \BibitemOpen
  \bibfield  {author} {\bibinfo {author} {\bibfnamefont {J.~A.}\ \bibnamefont
  {Ouassou}}, \bibinfo {author} {\bibfnamefont {A.}~\bibnamefont {Brataas}}, \
  and\ \bibinfo {author} {\bibfnamefont {J.}~\bibnamefont {Linder}},\
  }\bibfield  {title} {\bibinfo {title} {{dc Josephson Effect in
  Altermagnets}},\ }\href {\doibase 10.1103/PhysRevLett.131.076003} {\bibfield
  {journal} {\bibinfo  {journal} {Phys. Rev. Lett.}\ }\textbf {\bibinfo
  {volume} {131}},\ \bibinfo {pages} {076003} (\bibinfo {year}
  {2023})}\BibitemShut {NoStop}%
\bibitem [{\citenamefont {Sun}\ \emph {et~al.}(2023)\citenamefont {Sun},
  \citenamefont {Brataas},\ and\ \citenamefont {Linder}}]{Papaj2023}%
  \BibitemOpen
  \bibfield  {author} {\bibinfo {author} {\bibfnamefont {C.}~\bibnamefont
  {Sun}}, \bibinfo {author} {\bibfnamefont {A.}~\bibnamefont {Brataas}}, \ and\
  \bibinfo {author} {\bibfnamefont {J.}~\bibnamefont {Linder}},\ }\bibfield
  {title} {\bibinfo {title} {{Andreev reflection in altermagnets}},\ }\href
  {\doibase 10.1103/PhysRevB.108.054511} {\bibfield  {journal} {\bibinfo
  {journal} {Phys. Rev. B}\ }\textbf {\bibinfo {volume} {108}},\ \bibinfo
  {pages} {054511} (\bibinfo {year} {2023})}\BibitemShut {NoStop}%
\bibitem [{\citenamefont {Zhang}\ \emph {et~al.}(2024)\citenamefont {Zhang},
  \citenamefont {Hu},\ and\ \citenamefont {Neupert}}]{Neupert2024}%
  \BibitemOpen
  \bibfield  {author} {\bibinfo {author} {\bibfnamefont {S.-B.}\ \bibnamefont
  {Zhang}}, \bibinfo {author} {\bibfnamefont {L.-H.}\ \bibnamefont {Hu}}, \
  and\ \bibinfo {author} {\bibfnamefont {T.}~\bibnamefont {Neupert}},\
  }\bibfield  {title} {\bibinfo {title} {{Finite-momentum Cooper pairing in
  proximitized altermagnets}},\ }\href {\doibase 10.1038/s41467-024-45951-3}
  {\bibfield  {journal} {\bibinfo  {journal} {Nat. Commun.}\ }\textbf {\bibinfo
  {volume} {15}},\ \bibinfo {pages} {1801} (\bibinfo {year}
  {2024})}\BibitemShut {NoStop}%
\bibitem [{\citenamefont {Zhu}\ \emph {et~al.}(2023)\citenamefont {Zhu},
  \citenamefont {Zhuang}, \citenamefont {Wu},\ and\ \citenamefont
  {Yan}}]{Zhu2023}%
  \BibitemOpen
  \bibfield  {author} {\bibinfo {author} {\bibfnamefont {D.}~\bibnamefont
  {Zhu}}, \bibinfo {author} {\bibfnamefont {Z.-Y.}\ \bibnamefont {Zhuang}},
  \bibinfo {author} {\bibfnamefont {Z.}~\bibnamefont {Wu}}, \ and\ \bibinfo
  {author} {\bibfnamefont {Z.}~\bibnamefont {Yan}},\ }\bibfield  {title}
  {\bibinfo {title} {Topological superconductivity in two-dimensional
  altermagnetic metals},\ }\href
  {https://link.aps.org/doi/10.1103/PhysRevB.108.184505} {\bibfield  {journal}
  {\bibinfo  {journal} {Phys. Rev. B}\ }\textbf {\bibinfo {volume} {108}},\
  \bibinfo {pages} {184505} (\bibinfo {year} {2023})}\BibitemShut {NoStop}%
\bibitem [{\citenamefont {Brekke}\ \emph {et~al.}(2023)\citenamefont {Brekke},
  \citenamefont {Brataas},\ and\ \citenamefont {Sudb\o{}}}]{Sudbo2023}%
  \BibitemOpen
  \bibfield  {author} {\bibinfo {author} {\bibfnamefont {B.}~\bibnamefont
  {Brekke}}, \bibinfo {author} {\bibfnamefont {A.}~\bibnamefont {Brataas}}, \
  and\ \bibinfo {author} {\bibfnamefont {A.}~\bibnamefont {Sudb\o{}}},\
  }\bibfield  {title} {\bibinfo {title} {{Two-dimensional altermagnets:
  Superconductivity in a minimal microscopic model}},\ }\href {\doibase
  10.1103/PhysRevB.108.224421} {\bibfield  {journal} {\bibinfo  {journal}
  {Phys. Rev. B}\ }\textbf {\bibinfo {volume} {108}},\ \bibinfo {pages}
  {224421} (\bibinfo {year} {2023})}\BibitemShut {NoStop}%
\bibitem [{\citenamefont {Chakraborty}\ and\ \citenamefont
  {Black-Schaffer}(2024{\natexlab{a}})}]{Chakraborty2024}%
  \BibitemOpen
  \bibfield  {author} {\bibinfo {author} {\bibfnamefont {D.}~\bibnamefont
  {Chakraborty}}\ and\ \bibinfo {author} {\bibfnamefont {A.~M.}\ \bibnamefont
  {Black-Schaffer}},\ }\bibfield  {title} {\bibinfo {title} {{Zero-field
  finite-momentum and field-induced superconductivity in altermagnets}},\
  }\href {\doibase 10.1103/PhysRevB.110.L060508} {\bibfield  {journal}
  {\bibinfo  {journal} {Phys. Rev. B}\ }\textbf {\bibinfo {volume} {110}},\
  \bibinfo {pages} {L060508} (\bibinfo {year}
  {2024}{\natexlab{a}})}\BibitemShut {NoStop}%
\bibitem [{\citenamefont {Hong}\ \emph {et~al.}(2024)\citenamefont {Hong},
  \citenamefont {Park},\ and\ \citenamefont {Kim}}]{Hong2024}%
  \BibitemOpen
  \bibfield  {author} {\bibinfo {author} {\bibfnamefont {S.}~\bibnamefont
  {Hong}}, \bibinfo {author} {\bibfnamefont {M.~J.}\ \bibnamefont {Park}}, \
  and\ \bibinfo {author} {\bibfnamefont {K.-M.}\ \bibnamefont {Kim}},\
  }\bibfield  {title} {\bibinfo {title} {{Unconventional p-wave and
  finite-momentum superconductivity induced by altermagnetism through the
  formation of Bogoliubov Fermi surface}},\ }\href
  {https://arxiv.org/abs/2407.02059} {\bibfield  {journal} {\bibinfo  {journal}
  {arXiv:2407.02059}\ } (\bibinfo {year} {2024})}\BibitemShut {NoStop}%
\bibitem [{\citenamefont {Chakraborty}\ and\ \citenamefont
  {Black-Schaffer}(2024{\natexlab{b}})}]{Chakraborty-arXiv(2024)}%
  \BibitemOpen
  \bibfield  {author} {\bibinfo {author} {\bibfnamefont {D.}~\bibnamefont
  {Chakraborty}}\ and\ \bibinfo {author} {\bibfnamefont {A.~M.}\ \bibnamefont
  {Black-Schaffer}},\ }\bibfield  {title} {\bibinfo {title} {{Constraints on
  superconducting pairing in altermagnets}},\ }\href
  {https://arxiv.org/abs/2408.03999} {\bibfield  {journal} {\bibinfo  {journal}
  {arXiv:2408.03999}\ } (\bibinfo {year} {2024}{\natexlab{b}})}\BibitemShut
  {NoStop}%
\bibitem [{\citenamefont {Banerjee}\ and\ \citenamefont
  {Scheurer}(2024)}]{Scheurer-PRB(2024)}%
  \BibitemOpen
  \bibfield  {author} {\bibinfo {author} {\bibfnamefont {S.}~\bibnamefont
  {Banerjee}}\ and\ \bibinfo {author} {\bibfnamefont {M.~S.}\ \bibnamefont
  {Scheurer}},\ }\bibfield  {title} {\bibinfo {title} {{Altermagnetic
  superconducting diode effect}},\ }\href {\doibase
  10.1103/PhysRevB.110.024503} {\bibfield  {journal} {\bibinfo  {journal}
  {Phys. Rev. B}\ }\textbf {\bibinfo {volume} {110}},\ \bibinfo {pages}
  {024503} (\bibinfo {year} {2024})}\BibitemShut {NoStop}%
\bibitem [{\citenamefont {Sim}\ and\ \citenamefont
  {Knolle}(2024)}]{Knolle-arXiv(2024)}%
  \BibitemOpen
  \bibfield  {author} {\bibinfo {author} {\bibfnamefont {G.}~\bibnamefont
  {Sim}}\ and\ \bibinfo {author} {\bibfnamefont {J.}~\bibnamefont {Knolle}},\
  }\bibfield  {title} {\bibinfo {title} {{Pair Density Waves and Supercurrent
  Diode Effect in Altermagnets}},\ }\href {https://arxiv.org/abs/2407.01513}
  {\bibfield  {journal} {\bibinfo  {journal} {arXiv:2407.01513}\ } (\bibinfo
  {year} {2024})}\BibitemShut {NoStop}%
\bibitem [{\citenamefont {Li}\ \emph {et~al.}(2024)\citenamefont {Li},
  \citenamefont {Hu}, \citenamefont {Li}, \citenamefont {Wang}, \citenamefont
  {Chen}, \citenamefont {Thiagarajan}, \citenamefont {Leandersson},
  \citenamefont {Polley}, \citenamefont {Kim}, \citenamefont {Liu},
  \citenamefont {Fulga}, \citenamefont {Vergniory}, \citenamefont {Janson},
  \citenamefont {Tjernberg},\ and\ \citenamefont {van~den
  Brink}}]{vanderBrink2024}%
  \BibitemOpen
  \bibfield  {author} {\bibinfo {author} {\bibfnamefont {C.}~\bibnamefont
  {Li}}, \bibinfo {author} {\bibfnamefont {M.}~\bibnamefont {Hu}}, \bibinfo
  {author} {\bibfnamefont {Z.}~\bibnamefont {Li}}, \bibinfo {author}
  {\bibfnamefont {Y.}~\bibnamefont {Wang}}, \bibinfo {author} {\bibfnamefont
  {W.}~\bibnamefont {Chen}}, \bibinfo {author} {\bibfnamefont {B.}~\bibnamefont
  {Thiagarajan}}, \bibinfo {author} {\bibfnamefont {M.}~\bibnamefont
  {Leandersson}}, \bibinfo {author} {\bibfnamefont {C.}~\bibnamefont {Polley}},
  \bibinfo {author} {\bibfnamefont {T.}~\bibnamefont {Kim}}, \bibinfo {author}
  {\bibfnamefont {H.}~\bibnamefont {Liu}}, \bibinfo {author} {\bibfnamefont
  {C.}~\bibnamefont {Fulga}}, \bibinfo {author} {\bibfnamefont {M.~G.}\
  \bibnamefont {Vergniory}}, \bibinfo {author} {\bibfnamefont {O.}~\bibnamefont
  {Janson}}, \bibinfo {author} {\bibfnamefont {O.}~\bibnamefont {Tjernberg}}, \
  and\ \bibinfo {author} {\bibfnamefont {J.}~\bibnamefont {van~den Brink}},\
  }\bibfield  {title} {\bibinfo {title} {{Topological Weyl Altermagnetism in
  CrSb}},\ }\href {https://arxiv.org/abs/2405.14777} {\bibfield  {journal}
  {\bibinfo  {journal} {arXiv:2405.14777}\ } (\bibinfo {year}
  {2024})}\BibitemShut {NoStop}%
\bibitem [{\citenamefont {Lu}\ \emph {et~al.}(2024)\citenamefont {Lu},
  \citenamefont {Feng}, \citenamefont {Wang}, \citenamefont {Chen},
  \citenamefont {Lin}, \citenamefont {Liang}, \citenamefont {Liu},
  \citenamefont {Feng}, \citenamefont {Yamagami}, \citenamefont {Liu},
  \citenamefont {Felser}, \citenamefont {Wu},\ and\ \citenamefont
  {Ma}}]{Ma-arXiv(2024)}%
  \BibitemOpen
  \bibfield  {author} {\bibinfo {author} {\bibfnamefont {W.}~\bibnamefont
  {Lu}}, \bibinfo {author} {\bibfnamefont {S.}~\bibnamefont {Feng}}, \bibinfo
  {author} {\bibfnamefont {Y.}~\bibnamefont {Wang}}, \bibinfo {author}
  {\bibfnamefont {D.}~\bibnamefont {Chen}}, \bibinfo {author} {\bibfnamefont
  {Z.}~\bibnamefont {Lin}}, \bibinfo {author} {\bibfnamefont {X.}~\bibnamefont
  {Liang}}, \bibinfo {author} {\bibfnamefont {S.}~\bibnamefont {Liu}}, \bibinfo
  {author} {\bibfnamefont {W.}~\bibnamefont {Feng}}, \bibinfo {author}
  {\bibfnamefont {K.}~\bibnamefont {Yamagami}}, \bibinfo {author}
  {\bibfnamefont {J.}~\bibnamefont {Liu}}, \bibinfo {author} {\bibfnamefont
  {C.}~\bibnamefont {Felser}}, \bibinfo {author} {\bibfnamefont
  {Q.}~\bibnamefont {Wu}}, \ and\ \bibinfo {author} {\bibfnamefont
  {J.}~\bibnamefont {Ma}},\ }\bibfield  {title} {\bibinfo {title} {{Observation
  of surface Fermi arcs in altermagnetic Weyl semimetal CrSb}},\ }\href
  {https://arxiv.org/abs/2407.13497} {\bibfield  {journal} {\bibinfo  {journal}
  {arXiv:2407.13497}\ } (\bibinfo {year} {2024})}\BibitemShut {NoStop}%
\bibitem [{\citenamefont {Fernandes}\ \emph {et~al.}(2024)\citenamefont
  {Fernandes}, \citenamefont {de~Carvalho}, \citenamefont {Birol},\ and\
  \citenamefont {Pereira}}]{Fernandes-PRB(2024)}%
  \BibitemOpen
  \bibfield  {author} {\bibinfo {author} {\bibfnamefont {R.~M.}\ \bibnamefont
  {Fernandes}}, \bibinfo {author} {\bibfnamefont {V.~S.}\ \bibnamefont
  {de~Carvalho}}, \bibinfo {author} {\bibfnamefont {T.}~\bibnamefont {Birol}},
  \ and\ \bibinfo {author} {\bibfnamefont {R.~G.}\ \bibnamefont {Pereira}},\
  }\bibfield  {title} {\bibinfo {title} {{Topological transition from nodal to
  nodeless Zeeman splitting in altermagnets}},\ }\href {\doibase
  10.1103/PhysRevB.109.024404} {\bibfield  {journal} {\bibinfo  {journal}
  {Phys. Rev. B}\ }\textbf {\bibinfo {volume} {109}},\ \bibinfo {pages}
  {024404} (\bibinfo {year} {2024})}\BibitemShut {NoStop}%
\bibitem [{\citenamefont {Antonenko}\ \emph {et~al.}(2024)\citenamefont
  {Antonenko}, \citenamefont {Fernandes},\ and\ \citenamefont
  {Venderbos}}]{Venderbos-arXiv(2024)}%
  \BibitemOpen
  \bibfield  {author} {\bibinfo {author} {\bibfnamefont {D.~S.}\ \bibnamefont
  {Antonenko}}, \bibinfo {author} {\bibfnamefont {R.~M.}\ \bibnamefont
  {Fernandes}}, \ and\ \bibinfo {author} {\bibfnamefont {J.~W.~F.}\
  \bibnamefont {Venderbos}},\ }\bibfield  {title} {\bibinfo {title} {{Mirror
  Chern Bands and Weyl Nodal Loops in Altermagnets}},\ }\href
  {https://arxiv.org/abs/2402.10201} {\bibfield  {journal} {\bibinfo  {journal}
  {arXiv:2402.10201}\ } (\bibinfo {year} {2024})}\BibitemShut {NoStop}%
\bibitem [{\citenamefont {Agterberg}\ \emph {et~al.}(2017)\citenamefont
  {Agterberg}, \citenamefont {Brydon},\ and\ \citenamefont
  {Timm}}]{Agterberg-PRL(2017)}%
  \BibitemOpen
  \bibfield  {author} {\bibinfo {author} {\bibfnamefont {D.~F.}\ \bibnamefont
  {Agterberg}}, \bibinfo {author} {\bibfnamefont {P.~M.~R.}\ \bibnamefont
  {Brydon}}, \ and\ \bibinfo {author} {\bibfnamefont {C.}~\bibnamefont
  {Timm}},\ }\bibfield  {title} {\bibinfo {title} {{Bogoliubov Fermi Surfaces
  in Superconductors with Broken Time-Reversal Symmetry}},\ }\href {\doibase
  10.1103/PhysRevLett.118.127001} {\bibfield  {journal} {\bibinfo  {journal}
  {Phys. Rev. Lett.}\ }\textbf {\bibinfo {volume} {118}},\ \bibinfo {pages}
  {127001} (\bibinfo {year} {2017})}\BibitemShut {NoStop}%
\bibitem [{\citenamefont {Brydon}\ \emph {et~al.}(2018)\citenamefont {Brydon},
  \citenamefont {Agterberg}, \citenamefont {Menke},\ and\ \citenamefont
  {Timm}}]{Brydon-PRB(2018)}%
  \BibitemOpen
  \bibfield  {author} {\bibinfo {author} {\bibfnamefont {P.~M.~R.}\
  \bibnamefont {Brydon}}, \bibinfo {author} {\bibfnamefont {D.~F.}\
  \bibnamefont {Agterberg}}, \bibinfo {author} {\bibfnamefont {H.}~\bibnamefont
  {Menke}}, \ and\ \bibinfo {author} {\bibfnamefont {C.}~\bibnamefont {Timm}},\
  }\bibfield  {title} {\bibinfo {title} {{Bogoliubov Fermi surfaces: General
  theory, magnetic order, and topology}},\ }\href {\doibase
  10.1103/PhysRevB.98.224509} {\bibfield  {journal} {\bibinfo  {journal} {Phys.
  Rev. B}\ }\textbf {\bibinfo {volume} {98}},\ \bibinfo {pages} {224509}
  (\bibinfo {year} {2018})}\BibitemShut {NoStop}%
\bibitem [{\citenamefont {Chiu}\ and\ \citenamefont
  {Schnyder}(2014)}]{Schnyder2014}%
  \BibitemOpen
  \bibfield  {author} {\bibinfo {author} {\bibfnamefont {C.-K.}\ \bibnamefont
  {Chiu}}\ and\ \bibinfo {author} {\bibfnamefont {A.~P.}\ \bibnamefont
  {Schnyder}},\ }\bibfield  {title} {\bibinfo {title} {{Classification of
  reflection-symmetry-protected topological semimetals and nodal
  superconductors}},\ }\href {\doibase 10.1103/PhysRevB.90.205136} {\bibfield
  {journal} {\bibinfo  {journal} {Phys. Rev. B}\ }\textbf {\bibinfo {volume}
  {90}},\ \bibinfo {pages} {205136} (\bibinfo {year} {2014})}\BibitemShut
  {NoStop}%
\bibitem [{\citenamefont {Chiu}\ \emph {et~al.}(2016)\citenamefont {Chiu},
  \citenamefont {Teo}, \citenamefont {Schnyder},\ and\ \citenamefont
  {Ryu}}]{Chiu2015}%
  \BibitemOpen
  \bibfield  {author} {\bibinfo {author} {\bibfnamefont {C.-K.}\ \bibnamefont
  {Chiu}}, \bibinfo {author} {\bibfnamefont {J.~C.~Y.}\ \bibnamefont {Teo}},
  \bibinfo {author} {\bibfnamefont {A.~P.}\ \bibnamefont {Schnyder}}, \ and\
  \bibinfo {author} {\bibfnamefont {S.}~\bibnamefont {Ryu}},\ }\bibfield
  {title} {\bibinfo {title} {Classification of topological quantum matter with
  symmetries},\ }\href {\doibase 10.1103/RevModPhys.88.035005} {\bibfield
  {journal} {\bibinfo  {journal} {Rev. Mod. Phys.}\ }\textbf {\bibinfo {volume}
  {88}},\ \bibinfo {pages} {035005} (\bibinfo {year} {2016})}\BibitemShut
  {NoStop}%
\bibitem [{\citenamefont {Soluyanov}\ \emph {et~al.}(2015)\citenamefont
  {Soluyanov}, \citenamefont {Gresch}, \citenamefont {Wang}, \citenamefont
  {Wu}, \citenamefont {Troyer}, \citenamefont {Dai},\ and\ \citenamefont
  {Bernevig}}]{Bernevig2015}%
  \BibitemOpen
  \bibfield  {author} {\bibinfo {author} {\bibfnamefont {A.~A.}\ \bibnamefont
  {Soluyanov}}, \bibinfo {author} {\bibfnamefont {D.}~\bibnamefont {Gresch}},
  \bibinfo {author} {\bibfnamefont {Z.}~\bibnamefont {Wang}}, \bibinfo {author}
  {\bibfnamefont {Q.}~\bibnamefont {Wu}}, \bibinfo {author} {\bibfnamefont
  {M.}~\bibnamefont {Troyer}}, \bibinfo {author} {\bibfnamefont
  {X.}~\bibnamefont {Dai}}, \ and\ \bibinfo {author} {\bibfnamefont {B.~A.}\
  \bibnamefont {Bernevig}},\ }\bibfield  {title} {\bibinfo {title} {{Type-II
  Weyl semimetals}},\ }\href {\doibase 10.1038/nature15768} {\bibfield
  {journal} {\bibinfo  {journal} {Nature (London)}\ }\textbf {\bibinfo {volume}
  {527}},\ \bibinfo {pages} {495} (\bibinfo {year} {2015})}\BibitemShut
  {NoStop}%
\bibitem [{\citenamefont {Xu}\ \emph {et~al.}(2015)\citenamefont {Xu},
  \citenamefont {Zhang},\ and\ \citenamefont {Zhang}}]{Zhang-PRL(2015)}%
  \BibitemOpen
  \bibfield  {author} {\bibinfo {author} {\bibfnamefont {Y.}~\bibnamefont
  {Xu}}, \bibinfo {author} {\bibfnamefont {F.}~\bibnamefont {Zhang}}, \ and\
  \bibinfo {author} {\bibfnamefont {C.}~\bibnamefont {Zhang}},\ }\bibfield
  {title} {\bibinfo {title} {{Structured Weyl Points in Spin-Orbit Coupled
  Fermionic Superfluids}},\ }\href {\doibase 10.1103/PhysRevLett.115.265304}
  {\bibfield  {journal} {\bibinfo  {journal} {Phys. Rev. Lett.}\ }\textbf
  {\bibinfo {volume} {115}},\ \bibinfo {pages} {265304} (\bibinfo {year}
  {2015})}\BibitemShut {NoStop}%
\bibitem [{\citenamefont {Coleman}(2015)}]{Coleman-CUP(2015)}%
  \BibitemOpen
  \bibfield  {author} {\bibinfo {author} {\bibfnamefont {P.}~\bibnamefont
  {Coleman}},\ }\href@noop {} {\emph {\bibinfo {title} {{Introduction to
  Many-Body Physics}}}}\ (\bibinfo  {publisher} {Cambridge University Press},\
  \bibinfo {address} {Cambridge},\ \bibinfo {year} {2015})\BibitemShut
  {NoStop}%
\bibitem [{Sup()}]{Suppl_Mat}%
  \BibitemOpen
  \href@noop {} {}\bibinfo {note} {See \textcolor{blue}{Supplemental Material}
  at [\emph{URL will be inserted by publisher}] for further information on the
  derivation of the superconducting gap equations, the stability analysis of
  the superconducting nodal manifolds, and the calculation of the coefficients
  of the Ginzburg-Landau free energy.}\BibitemShut {Stop}%
\bibitem [{\citenamefont {Frigeri}\ \emph {et~al.}(2004)\citenamefont
  {Frigeri}, \citenamefont {Agterberg}, \citenamefont {Koga},\ and\
  \citenamefont {Sigrist}}]{Sigrist-PRL(2004)}%
  \BibitemOpen
  \bibfield  {author} {\bibinfo {author} {\bibfnamefont {P.~A.}\ \bibnamefont
  {Frigeri}}, \bibinfo {author} {\bibfnamefont {D.~F.}\ \bibnamefont
  {Agterberg}}, \bibinfo {author} {\bibfnamefont {A.}~\bibnamefont {Koga}}, \
  and\ \bibinfo {author} {\bibfnamefont {M.}~\bibnamefont {Sigrist}},\
  }\bibfield  {title} {\bibinfo {title} {{Superconductivity without Inversion
  Symmetry: MnSi versus
  ${\mathrm{C}\mathrm{e}\mathrm{P}\mathrm{t}}_{3}\mathrm{S}\mathrm{i}$}},\
  }\href {\doibase 10.1103/PhysRevLett.92.097001} {\bibfield  {journal}
  {\bibinfo  {journal} {Phys. Rev. Lett.}\ }\textbf {\bibinfo {volume} {92}},\
  \bibinfo {pages} {097001} (\bibinfo {year} {2004})}\BibitemShut {NoStop}%
\bibitem [{Note1()}]{Note1}%
  \BibitemOpen
  \bibinfo {note} {The AM form factors $\protect \bm {g}_{\protect \mathbf
  {k}}$ used in the numerical calculations presented in this work are
  normalized at the Fermi surface.}\BibitemShut {Stop}%
\bibitem [{\citenamefont {Anderson}(1959)}]{Anderson-JPCS(1959)}%
  \BibitemOpen
  \bibfield  {author} {\bibinfo {author} {\bibfnamefont {P.}~\bibnamefont
  {Anderson}},\ }\bibfield  {title} {\bibinfo {title} {{Theory of dirty
  superconductors}},\ }\href {\doibase
  https://doi.org/10.1016/0022-3697(59)90036-8} {\bibfield  {journal} {\bibinfo
   {journal} {J. Phys. Chem. Solids}\ }\textbf {\bibinfo {volume} {11}},\
  \bibinfo {pages} {26} (\bibinfo {year} {1959})}\BibitemShut {NoStop}%
\bibitem [{\citenamefont {Kobayashi}\ \emph {et~al.}(2014)\citenamefont
  {Kobayashi}, \citenamefont {Shiozaki}, \citenamefont {Tanaka},\ and\
  \citenamefont {Sato}}]{Kobayashi-PRB(2014)}%
  \BibitemOpen
  \bibfield  {author} {\bibinfo {author} {\bibfnamefont {S.}~\bibnamefont
  {Kobayashi}}, \bibinfo {author} {\bibfnamefont {K.}~\bibnamefont {Shiozaki}},
  \bibinfo {author} {\bibfnamefont {Y.}~\bibnamefont {Tanaka}}, \ and\ \bibinfo
  {author} {\bibfnamefont {M.}~\bibnamefont {Sato}},\ }\bibfield  {title}
  {\bibinfo {title} {{Topological Blount's theorem of odd-parity
  superconductors}},\ }\href {\doibase 10.1103/PhysRevB.90.024516} {\bibfield
  {journal} {\bibinfo  {journal} {Phys. Rev. B}\ }\textbf {\bibinfo {volume}
  {90}},\ \bibinfo {pages} {024516} (\bibinfo {year} {2014})}\BibitemShut
  {NoStop}%
\bibitem [{\citenamefont {Zhao}\ \emph {et~al.}(2016)\citenamefont {Zhao},
  \citenamefont {Schnyder},\ and\ \citenamefont {Wang}}]{Zhao-PRL(2016)}%
  \BibitemOpen
  \bibfield  {author} {\bibinfo {author} {\bibfnamefont {Y.~X.}\ \bibnamefont
  {Zhao}}, \bibinfo {author} {\bibfnamefont {A.~P.}\ \bibnamefont {Schnyder}},
  \ and\ \bibinfo {author} {\bibfnamefont {Z.~D.}\ \bibnamefont {Wang}},\
  }\bibfield  {title} {\bibinfo {title} {{Unified Theory of $PT$ and $CP$
  Invariant Topological Metals and Nodal Superconductors}},\ }\href {\doibase
  10.1103/PhysRevLett.116.156402} {\bibfield  {journal} {\bibinfo  {journal}
  {Phys. Rev. Lett.}\ }\textbf {\bibinfo {volume} {116}},\ \bibinfo {pages}
  {156402} (\bibinfo {year} {2016})}\BibitemShut {NoStop}%
\bibitem [{\citenamefont {Fernandes}\ \emph {et~al.}(2012)\citenamefont
  {Fernandes}, \citenamefont {Chubukov}, \citenamefont {Knolle}, \citenamefont
  {Eremin},\ and\ \citenamefont {Schmalian}}]{Fernandes-PRB(2012)}%
  \BibitemOpen
  \bibfield  {author} {\bibinfo {author} {\bibfnamefont {R.~M.}\ \bibnamefont
  {Fernandes}}, \bibinfo {author} {\bibfnamefont {A.~V.}\ \bibnamefont
  {Chubukov}}, \bibinfo {author} {\bibfnamefont {J.}~\bibnamefont {Knolle}},
  \bibinfo {author} {\bibfnamefont {I.}~\bibnamefont {Eremin}}, \ and\ \bibinfo
  {author} {\bibfnamefont {J.}~\bibnamefont {Schmalian}},\ }\bibfield  {title}
  {\bibinfo {title} {{Preemptive nematic order, pseudogap, and orbital order in
  the iron pnictides}},\ }\href {\doibase 10.1103/PhysRevB.85.024534}
  {\bibfield  {journal} {\bibinfo  {journal} {Phys. Rev. B}\ }\textbf {\bibinfo
  {volume} {85}},\ \bibinfo {pages} {024534} (\bibinfo {year}
  {2012})}\BibitemShut {NoStop}%
\bibitem [{\citenamefont {Hinojosa}\ \emph {et~al.}(2014)\citenamefont
  {Hinojosa}, \citenamefont {Fernandes},\ and\ \citenamefont
  {Chubukov}}]{Hinojosa-PRL(2014)}%
  \BibitemOpen
  \bibfield  {author} {\bibinfo {author} {\bibfnamefont {A.}~\bibnamefont
  {Hinojosa}}, \bibinfo {author} {\bibfnamefont {R.~M.}\ \bibnamefont
  {Fernandes}}, \ and\ \bibinfo {author} {\bibfnamefont {A.~V.}\ \bibnamefont
  {Chubukov}},\ }\bibfield  {title} {\bibinfo {title} {{Time-Reversal Symmetry
  Breaking Superconductivity in the Coexistence Phase with Magnetism in Fe
  Pnictides}},\ }\href {\doibase 10.1103/PhysRevLett.113.167001} {\bibfield
  {journal} {\bibinfo  {journal} {Phys. Rev. Lett.}\ }\textbf {\bibinfo
  {volume} {113}},\ \bibinfo {pages} {167001} (\bibinfo {year}
  {2014})}\BibitemShut {NoStop}%
\bibitem [{\citenamefont {Altland}\ and\ \citenamefont
  {Simons}(2010)}]{Altland-CUP(2010)}%
  \BibitemOpen
  \bibfield  {author} {\bibinfo {author} {\bibfnamefont {A.}~\bibnamefont
  {Altland}}\ and\ \bibinfo {author} {\bibfnamefont {B.~D.}\ \bibnamefont
  {Simons}},\ }\href@noop {} {\emph {\bibinfo {title} {{Condensed Matter Field
  Theory}}}},\ \bibinfo {edition} {2nd}\ ed.\ (\bibinfo  {publisher} {Cambridge
  University Press},\ \bibinfo {address} {Cambridge},\ \bibinfo {year}
  {2010})\BibitemShut {NoStop}%
\bibitem [{\citenamefont {Lee}\ \emph {et~al.}(2009)\citenamefont {Lee},
  \citenamefont {Zhang},\ and\ \citenamefont {Wu}}]{Wu-PRL(2009)}%
  \BibitemOpen
  \bibfield  {author} {\bibinfo {author} {\bibfnamefont {W.-C.}\ \bibnamefont
  {Lee}}, \bibinfo {author} {\bibfnamefont {S.-C.}\ \bibnamefont {Zhang}}, \
  and\ \bibinfo {author} {\bibfnamefont {C.}~\bibnamefont {Wu}},\ }\bibfield
  {title} {\bibinfo {title} {{Pairing State with a Time-Reversal Symmetry
  Breaking in FeAs-Based Superconductors}},\ }\href {\doibase
  10.1103/PhysRevLett.102.217002} {\bibfield  {journal} {\bibinfo  {journal}
  {Phys. Rev. Lett.}\ }\textbf {\bibinfo {volume} {102}},\ \bibinfo {pages}
  {217002} (\bibinfo {year} {2009})}\BibitemShut {NoStop}%
\bibitem [{\citenamefont {Wu}\ and\ \citenamefont
  {Hirsch}(2010)}]{Wu-PRB(2010)}%
  \BibitemOpen
  \bibfield  {author} {\bibinfo {author} {\bibfnamefont {C.}~\bibnamefont
  {Wu}}\ and\ \bibinfo {author} {\bibfnamefont {J.~E.}\ \bibnamefont
  {Hirsch}},\ }\bibfield  {title} {\bibinfo {title} {{Mixed triplet and singlet
  pairing in ultracold multicomponent fermion systems with dipolar
  interactions}},\ }\href {\doibase 10.1103/PhysRevB.81.020508} {\bibfield
  {journal} {\bibinfo  {journal} {Phys. Rev. B}\ }\textbf {\bibinfo {volume}
  {81}},\ \bibinfo {pages} {020508} (\bibinfo {year} {2010})}\BibitemShut
  {NoStop}%
\bibitem [{\citenamefont {Maiti}\ and\ \citenamefont
  {Chubukov}(2013)}]{Chubukov-PRB(2013)}%
  \BibitemOpen
  \bibfield  {author} {\bibinfo {author} {\bibfnamefont {S.}~\bibnamefont
  {Maiti}}\ and\ \bibinfo {author} {\bibfnamefont {A.~V.}\ \bibnamefont
  {Chubukov}},\ }\bibfield  {title} {\bibinfo {title} {{$s+is$ state with
  broken time-reversal symmetry in Fe-based superconductors}},\ }\href
  {\doibase 10.1103/PhysRevB.87.144511} {\bibfield  {journal} {\bibinfo
  {journal} {Phys. Rev. B}\ }\textbf {\bibinfo {volume} {87}},\ \bibinfo
  {pages} {144511} (\bibinfo {year} {2013})}\BibitemShut {NoStop}%
\bibitem [{\citenamefont {Wang}\ and\ \citenamefont
  {Chubukov}(2014)}]{Chubukov-PRB(2014)}%
  \BibitemOpen
  \bibfield  {author} {\bibinfo {author} {\bibfnamefont {Y.}~\bibnamefont
  {Wang}}\ and\ \bibinfo {author} {\bibfnamefont {A.}~\bibnamefont
  {Chubukov}},\ }\bibfield  {title} {\bibinfo {title} {{Charge-density-wave
  order with momentum $(2Q,0)$ and $(0,2Q)$ within the spin-fermion model:
  Continuous and discrete symmetry breaking, preemptive composite order, and
  relation to pseudogap in hole-doped cuprates}},\ }\href {\doibase
  10.1103/PhysRevB.90.035149} {\bibfield  {journal} {\bibinfo  {journal} {Phys.
  Rev. B}\ }\textbf {\bibinfo {volume} {90}},\ \bibinfo {pages} {035149}
  (\bibinfo {year} {2014})}\BibitemShut {NoStop}%
\bibitem [{\citenamefont {Wang}\ and\ \citenamefont
  {Fu}(2017)}]{Wang-PRL(2017)}%
  \BibitemOpen
  \bibfield  {author} {\bibinfo {author} {\bibfnamefont {Y.}~\bibnamefont
  {Wang}}\ and\ \bibinfo {author} {\bibfnamefont {L.}~\bibnamefont {Fu}},\
  }\bibfield  {title} {\bibinfo {title} {{Topological Phase Transitions in
  Multicomponent Superconductors}},\ }\href {\doibase
  10.1103/PhysRevLett.119.187003} {\bibfield  {journal} {\bibinfo  {journal}
  {Phys. Rev. Lett.}\ }\textbf {\bibinfo {volume} {119}},\ \bibinfo {pages}
  {187003} (\bibinfo {year} {2017})}\BibitemShut {NoStop}%
\bibitem [{Note2()}]{Note2}%
  \BibitemOpen
  \bibinfo {note} {The $d_{xy}$-wave SC state was not taken into account
  because it has a lower critical temperature compared to the $s$-wave and
  $d_{x^2 - y^2}$-wave SC states, when the AM interaction is switched
  on.}\BibitemShut {Stop}%
\bibitem [{Note3()}]{Note3}%
  \BibitemOpen
  \bibinfo {note} {However, we cannot rule out the possibility that this
  picture could change with the inclusion of higher-order terms in the
  expansion of the GL free energy. The latter study is nevertheless beyond the
  scope of the present work.}\BibitemShut {Stop}%
\end{thebibliography}

%

\end{document}


\title{\textbf{\Large Supplemental Material:} \vspace{0.4cm} \\ ``Unconventional superconductivity in altermagnets with spin-orbit coupling"}

\author{Vanuildo S. de Carvalho}
\affiliation{Instituto de F\'{\i}sica, Universidade Federal de Goi\'as, 74.690-900, Goi\^ania-GO,
Brazil}
\author{Hermann Freire}
\affiliation{Instituto de F\'{\i}sica, Universidade Federal de Goi\'as, 74.690-900, Goi\^ania-GO,
Brazil}

\setcounter{equation}{0}
\setcounter{figure}{0}
\setcounter{table}{0}
\setcounter{page}{1}
\setcounter{section}{0}
\makeatletter
\renewcommand{\thesection}{S\Roman{section}}
\renewcommand{\theequation}{S\arabic{equation}}
\renewcommand{\thefigure}{S\arabic{figure}}
\renewcommand{\thetable}{S\Roman{table}}
\renewcommand{\bibnumfmt}[1]{[S#1]}
\renewcommand{\citenumfont}[1]{S#1}

\maketitle

In this Supplemental Material, we extend the analysis and also give detailed derivations of the results presented in the main text.

\section{Superconducting gap equations}
\noindent

The properties of three-dimensional (3D) altermagnets in the presence of spin-orbit coupling (SOC) \cite{Fernandes-PRB(2024)} and weak pairing interactions are described by the Hamiltonian $H = H_0 + H_{\mathrm{AM}} + H_{\mathrm{SC}}$:
\begin{align}
H_0 & = \sum_{\mathbf{k}, s} \xi_{\mathbf{k}} \psi^{\dagger}_{\mathbf{k}, s} \psi_{\mathbf{k}, s}, \\
H_{\mathrm{AM}} & = \lambda \sum_{\mathbf{k}, s, s'} \psi^{\dagger}_{\mathbf{k}, s}[\boldsymbol{g}_{\mathbf{k}} \cdot \hat{\boldsymbol{\sigma}}]_{s,s'} \psi_{\mathbf{k}, s'}, \label{AM_SMHam}\\
H_{\mathrm{SC}} & = \frac{1}{2 \mathcal{V}} \sum_{\substack{\mathbf{k}, \mathbf{k}', \mathbf{q} \\ s, s'}} V_{\mathbf{k}, \mathbf{k}'} \psi^{\dagger}_{\mathbf{k} + \mathbf{q}/2, s} \psi^{\dagger}_{- \mathbf{k} + \mathbf{q}/2, s'} \psi_{- \mathbf{k}' + \mathbf{q}/2, s'} \psi_{\mathbf{k}' + \mathbf{q}/2, s}. \label{SC_SMHam01}
\end{align}
Here, $\xi_{\mathbf{k}} = \frac{\mathbf{k}^2}{2m} - \mu$ represents the energy dispersion of electrons with spin projection $s \in \{\uparrow, \downarrow \}$, $\lambda$ is the altermagnetic (AM) interaction, $\hat{\boldsymbol{\sigma}} \equiv (\hat{\sigma}_x,\hat{\sigma}_y,\hat{\sigma}_z)$ refers to the vector of Pauli matrices, $\mathcal{V}$ denotes the volume of the system, and $V_{\mathbf{k}, \mathbf{k}'}$ describes the pairing interaction which favors equally spin-singlet and spin-triplet superconductivity. Moreover, $\boldsymbol{g}_{\mathbf{k}}$ denotes the form factor for 3D altermagnets in the presence of SOC (see Ref. \cite{Fernandes-PRB(2024)} for more details). Even though our main results do not rely on the actual form of $\boldsymbol{g}_{\mathbf{k}}$, in the explicit calculations presented here we will consider $\boldsymbol{g}_{\mathbf{k}} = k_y k_z \hat{\boldsymbol{x}} + k_x k_z \hat{\boldsymbol{y}} + k_x k_y \hat{\boldsymbol{z}}$, which describes AM systems whose order parameter transforms as the $B^{-}_{1g}$ irreducible representation (irrep) of the $D_{4h}$ point group.

The mean-field decoupling of Eq. \eqref{SC_SMHam01} yields:
\begin{align}
H_{\mathrm{SC}} & = \frac{1}{2 \mathcal{V}} \sum_{\substack{\mathbf{k}, \mathbf{k}', \mathbf{q} \\ s, s'}} V_{\mathbf{k}, \mathbf{k}'} [ \psi^{\dagger}_{\mathbf{k} + \mathbf{q}/2, s} \psi^{\dagger}_{- \mathbf{k} + \mathbf{q}/2, s'} \langle \psi_{- \mathbf{k}' + \mathbf{q}/2, s'} \psi_{\mathbf{k}' + \mathbf{q}/2, s}\rangle + \langle \psi^{\dagger}_{\mathbf{k} + \mathbf{q}/2, s} \psi^{\dagger}_{- \mathbf{k} + \mathbf{q}/2, s'} \rangle \psi_{- \mathbf{k}' + \mathbf{q}/2, s'} \psi_{\mathbf{k}' + \mathbf{q}/2, s} \nonumber \\
& - \langle \psi^{\dagger}_{\mathbf{k} + \mathbf{q}/2, s} \psi^{\dagger}_{- \mathbf{k} + \mathbf{q}/2, s'} \rangle \langle \psi_{- \mathbf{k}' + \mathbf{q}/2, s'} \psi_{\mathbf{k}' + \mathbf{q}/2, s}\rangle], \label{SC_SMHam02}
\end{align}
where $\langle \psi_{- \mathbf{k} + \mathbf{q}/2, s'} \psi_{\mathbf{k} + \mathbf{q}/2, s} \rangle \equiv \delta_{\mathbf{q}, \boldsymbol{0}} [\widehat{\mathscr{F}}(\mathbf{k}) i \hat{\sigma}_y]_{s, s'}
$ and $\widehat{\mathscr{F}}(\mathbf{k})$ denotes a matrix in spin space, which is obtained self-consistently. Consequently, Eq. \eqref{SC_SMHam02} can be rewritten as:
\begin{equation}\label{SC_SMHam03}
H_{\mathrm{SC}} = - \frac{1}{2} \sum_{\mathbf{k}} \{ \psi^{\dagger}_{\mathbf{k}} [\widehat{\Delta}(\mathbf{k}) i\hat{\sigma}_y] (\psi^{\dagger}_{- \mathbf{k}})^T + \psi^T_{- \mathbf{k}} [\widehat{\Delta}(\mathbf{k}) i\hat{\sigma}_y]^{\dagger} \psi_{\mathbf{k}} \} + \frac{1}{2} \sum_{\mathbf{k}} \operatorname{tr} [\widehat{\Delta}(\mathbf{k}) \cdot \widehat{\mathscr{F}}^{\dagger}(\mathbf{k})].
\end{equation}
In the above equation, we have defined $\psi_{\mathbf{k}} \equiv ( \psi_{\mathbf{k}, \uparrow}, \; \psi_{\mathbf{k}, \downarrow} )^T$ and $\widehat{\Delta}(\mathbf{k}) \equiv - \dfrac{1}{\mathcal{V}} \sum\limits_{\mathbf{k}'} V_{\mathbf{k}, \mathbf{k}'} \widehat{\mathscr{F}}(\mathbf{k}')$, such that $\widehat{\Delta}(\mathbf{k})$ refers to the superconducting (SC) order parameter.

After employing the Balian-Werthamer spinor $\Psi_{\mathbf{k}} \equiv (\psi_{\mathbf{k}, \uparrow}, \; \psi_{\mathbf{k}, \downarrow}, \; - \psi^{\dagger}_{-\mathbf{k}, \downarrow}, \; \psi^{\dagger}_{-\mathbf{k}, \uparrow})^T$, we find that the Bogoliubov-de Gennes (BdG) Hamiltonian related to $H = H_0 + H_{\mathrm{AM}} + H_{\mathrm{SC}}$ evaluates to
\begin{equation}\label{Eq_Int_Ham}
H_{\mathrm{BdG}} = \frac{1}{2} \sum_{\mathbf{k}} \Psi^{\dagger}_{\mathbf{k}} \begin{pmatrix} \widehat{\mathcal{H}}_0(\mathbf{k}) & \widehat{\Delta}(\mathbf{k}) \\ \widehat{\Delta}^{\dagger}(\mathbf{k}) & - \hat{\sigma}_y \widehat{\mathcal{H}}^{T}_0(- \mathbf{k}) \hat{\sigma}_y \end{pmatrix} \Psi_{\mathbf{k}} + \frac{1}{2} \sum_{\mathbf{k}} \operatorname{tr} [\widehat{\Delta}(\mathbf{k}) \cdot \widehat{\mathscr{F}}^{\dagger}(\mathbf{k})] + \frac{1}{2} \sum_{\mathbf{k}} \operatorname{tr} [\widehat{\mathcal{H}}_0(\mathbf{k})],
\end{equation}
where $\widehat{\mathcal{H}}_0(\mathbf{k}) \equiv \xi_{\mathbf{k}} \hat{\sigma}_0 + \lambda \boldsymbol{g}_{\mathbf{k}} \cdot \hat{\boldsymbol{\sigma}}$ is the lattice Hamiltonian for a pure AM phase. The last term in the r.h.s. of the above equation only gives rise to a constant and will be dropped henceforth. In addition, we parametrize $\widehat{\Delta}(\mathbf{k})$ according to
\begin{equation}
\widehat{\Delta}(\mathbf{k}) = \Delta_s(\mathbf{k}) \hat{\sigma}_0 + \boldsymbol{\Delta}_t(\mathbf{k}) \cdot \hat{\boldsymbol{\sigma}},
\end{equation}
where $\Delta_s(\mathbf{k}) = \vert \Delta_s \vert e^{i \theta_s} d_0(\mathbf{k})$ and $\boldsymbol{\Delta}_t(\mathbf{k}) = \vert \Delta_t \vert e^{i \theta_t} \boldsymbol{d}(\mathbf{k})$ refer to the singlet and triplet pairing amplitudes, respectively. Notice that the SC state breaks time-reversal symmetry provided the phase difference $\theta_s - \theta_t$ is non-trivial. In addition, as required by Fermi-Dirac statistics, $d_0(\mathbf{k}) = d_0(- \mathbf{k})$ and $\boldsymbol{d}(\mathbf{k}) = - \boldsymbol{d}(- \mathbf{k})$.

The action associated to $H_{\mathrm{BdG}}$ reads:
\begin{equation}\label{SM_Action01}
\mathcal{S}[\overline{\Psi}, \Psi; \widehat{\Delta}^\dagger, \widehat{\Delta}] = \frac{1}{2} \int^{\beta}_0 d\tau \sum_{\mathbf{k}} \overline{\Psi}_{\mathbf{k}}(\tau) \begin{pmatrix} \hat{\sigma}_0 \partial_{\tau} + \widehat{\mathcal{H}}_0(\mathbf{k}) & \widehat{\Delta}(\mathbf{k}) \\ \widehat{\Delta}^{\dagger}(\mathbf{k}) & \hat{\sigma}_0 \partial_{\tau} - \hat{\sigma}_y \widehat{\mathcal{H}}^{T}_0(- \mathbf{k}) \hat{\sigma}_y \end{pmatrix} \Psi_{\mathbf{k}}(\tau) + \frac{1}{2}\int^{\beta}_0 d\tau \sum_{\mathbf{k}} \operatorname{tr} [\widehat{\Delta}(\mathbf{k}) \cdot \widehat{\mathscr{F}}^{\dagger}(\mathbf{k})].
\end{equation}
After integrating out the fermionic fields in $\mathcal{S}[\overline{\Psi}, \Psi; \widehat{\Delta}^\dagger, \widehat{\Delta}]$, the effective action for the SC state with singlet and triplet correlations evaluates to:
\begin{equation}\label{SM_Action02}
\mathcal{S}_{\mathrm{eff}}[\widehat{\Delta}^\dagger, \widehat{\Delta}] = - \mathrm{Tr}\ln\left(\widehat{\mathcal{G}}^{-1}\right) + \frac{1}{2}\int^{\beta}_0 d\tau \sum_{\mathbf{k}} \mathrm{tr}[\widehat{\Delta}(\mathbf{k}) \cdot \widehat{\mathscr{F}}^\dagger(\mathbf{k})],
\end{equation}
where
\begin{equation}
\widehat{\mathcal{G}}^{-1} \equiv \begin{pmatrix} \widehat{G}^{-1}_{0, p} & \widehat{\Delta} \\ \widehat{\Delta}^{\dagger} & \widehat{G}^{-1}_{0, h} \end{pmatrix}, \hspace{0.5cm}
\widehat{G}^{-1}_{0, p} \equiv \hat{\sigma}_0\partial_{\tau} + \widehat{\mathcal{H}}_0(- i\boldsymbol{\nabla}), \hspace{0.5cm}
\widehat{G}^{-1}_{0, h} \equiv \hat{\sigma}_0\partial_{\tau} - \hat{\sigma}_y\widehat{\mathcal{H}}_0(i\boldsymbol{\nabla})\hat{\sigma}_y.
\end{equation}
The Green's functions $\widehat{G}_{0, p}$ and $\widehat{G}_{0, h}$ refer to the particle and hole propagators, respectively. The minimization of the action $\mathcal{S}_{\mathrm{eff}}[\widehat{\Delta}^\dagger, \widehat{\Delta}]$ with respect to $\widehat{\Delta}$ yields the following gap equation:
\begin{equation}\label{SM_GapEq_01}
\widehat{\Delta}^{\dagger}(\mathbf{k}) = \frac{1}{\beta \mathcal{V}} \sum_{n, \mathbf{k}'} V_{\mathbf{k}, \mathbf{k}'}\{ [\widehat{G}_{0, h}(\mathbf{k}', i\omega_n)]^{-1} - \widehat{\Delta}^{\dagger}(\mathbf{k}') \widehat{G}_{0, p}(\mathbf{k}', i\omega_n)\widehat{\Delta}(\mathbf{k}') \}^{-1} \widehat{\Delta}^{\dagger}(\mathbf{k}') \widehat{G}_{0, p}(\mathbf{k}', i\omega_n),
\end{equation}
where $[\widehat{G}_{0, p}(\mathbf{k}, i \omega_n)]^{-1} = - i \omega_n \hat{\sigma}_0 + \widehat{\mathcal{H}}_0(\mathbf{k})$ and $[\widehat{G}_{0, h}(\mathbf{k}, i \omega_n)]^{-1} = - i \omega_n \hat{\sigma}_0 - \hat{\sigma}_y \widehat{\mathcal{H}}^{T}_0(- \mathbf{k}) \hat{\sigma}_y$. 

Since our main interest here is to determine $T_c$, we linearize the gap equation \eqref{SM_GapEq_01} and then make the substitution $\widehat{\Delta}(\mathbf{k}) = \Delta_s(\mathbf{k}) \hat{\sigma}_0 + \boldsymbol{\Delta}_t(\mathbf{k}) \cdot \hat{\boldsymbol{\sigma}}$. To find the gap equations for $\Delta_s(\mathbf{k})$ and $\boldsymbol{\Delta}_t(\mathbf{k})$, we first rewrite the particle and hole propagators as
\begin{align}
\widehat{G}_{0, p}(\mathbf{k}, i \omega_n) & = G_{+}(\mathbf{k}, i \omega_n) \hat{\sigma}_0 + G_{-}(\mathbf{k}, i \omega_n) \widehat{\boldsymbol{g}}_{\mathbf{k}} \cdot \hat{\boldsymbol{\sigma}}, \label{SM_PartProp}\\
\widehat{G}_{0, h}(\mathbf{k}, i \omega_n) & = - G_{+}(- \mathbf{k}, - i \omega_n) \hat{\sigma}_0 + G_{-}(- \mathbf{k}, - i \omega_n) \widehat{\boldsymbol{g}}_{\mathbf{k}} \cdot \hat{\boldsymbol{\sigma}}, \label{SM_HoleProp}
\end{align}
where the functions on the r.h.s. of the above equations read:
\begin{equation}
G_{\pm}(\mathbf{k}, i \omega_n) = \frac{1}{2}\left[ \frac{1}{-i \omega_n + E_{+}(\mathbf{k})} \pm \frac{1}{-i \omega_n + E_{-}(\mathbf{k})} \right],
\end{equation}
with $E_{\pm}(\mathbf{k}) = \xi_{\mathbf{k}} \pm \lambda \| \boldsymbol{g}_{\mathbf{k}} \|$ and $\widehat{\boldsymbol{g}}_{\mathbf{k}} = \boldsymbol{g}_{\mathbf{k}}/\| \boldsymbol{g}_{\mathbf{k}}\|$. The energy dispersions $E_{\pm}(\mathbf{k})$ give rise to two Fermi-surface sheets, which are shown schematically in Fig. 1(b) of the main text. As explained in Ref. \cite{Fernandes-PRB(2024)} (see also the main text), they intersect at Weyl pinching points defined by the conditions: $\xi_{\mathbf{k}} = 0$ and $\| \boldsymbol{g}_{\mathbf{k}} \| = 0$. In terms of the propagators $G_{\pm}(\mathbf{k}, i \omega_n)$, the linearized gap equations for $\Delta_s(\mathbf{k})$ and $\boldsymbol{\Delta}_t(\mathbf{k})$ turn out to be:
\begin{align}
\Delta^{*}_s(\mathbf{k}) & = - \frac{1}{\beta \mathcal{V}} \sum_{n, \mathbf{k}'} V_{\mathbf{k}, \mathbf{k}'} [ (\widetilde{G}_{+}G_{+} - \widetilde{G}_{-}G_{-}) \Delta^{*}_s(\mathbf{k}') + (\widetilde{G}_{+}G_{-} - \widetilde{G}_{-}G_{+}) \boldsymbol{\Delta}^{*}_t(\mathbf{k}') \cdot \widehat{\boldsymbol{g}}_{\mathbf{k}'} ], \label{SM_GapEq_02}\\
\boldsymbol{\Delta}^{*}_t(\mathbf{k}) & = - \frac{1}{\beta \mathcal{V}} \sum_{n, \mathbf{k}'} V_{\mathbf{k}, \mathbf{k}'} \lbrace (\widetilde{G}_{+} G_{+} - \widetilde{G}_{-} G_{-}) \boldsymbol{\Delta}^{*}_t(\mathbf{k}') - 2 \widetilde{G}_{-} G_{-} [(\boldsymbol{\Delta}^{*}_t(\mathbf{k}') \cdot \widehat{\boldsymbol{g}}_{\mathbf{k}'}) \widehat{\boldsymbol{g}}_{\mathbf{k}'} - \boldsymbol{\Delta}^{*}_t(\mathbf{k}')] \nonumber \\
& + (\widetilde{G}_{+} G_{-} - \widetilde{G}_{-} G_{+})\Delta^{*}_s(\mathbf{k}') \widehat{\boldsymbol{g}}_{\mathbf{k}'} + i (\widetilde{G}_{+} G_{-} + \widetilde{G}_{-} G_{+}) \boldsymbol{\Delta}^{*}_t(\mathbf{k}') \times \widehat{\boldsymbol{g}}_{\mathbf{k}'} \rbrace, \label{SM_GapEq_03}
\end{align}
where we have set $\widetilde{G}_{a} G_{b} \equiv G_{a}(- \mathbf{k}, - i \omega_n) G_{b}(\mathbf{k}, i \omega_n)$ with $a,b = \pm$. 

Henceforth, we follow the standard weak-coupling approach for superconductivity \cite{Coleman-CUP(2015)}, in which the pairing interaction $V_{\mathbf{k}, \mathbf{k}'}$ is assumed to be finite and attractive within an energy cutoff $\epsilon_c$. Besides, we restrict the momentum dependence of $V_{\mathbf{k}, \mathbf{k}'}$ to its angular coordinates. As a result, the evaluation of the propagators in Eqs. \eqref{SM_GapEq_02}-\eqref{SM_GapEq_03} is obtained by setting the density of states to its Fermi-surface value $N_F$ and then integrating over the momenta transversal to the Fermi surface. After computing the remaining Matsubara sums, we find:
\begin{align}
\frac{1}{\beta} \sum_{\vert \omega_n \vert < \epsilon_c} \int^{\infty}_{- \infty} d\xi G_{+}(- \mathbf{k}, - i\omega_n) G_{+}(\mathbf{k}, i\omega_n) & = \ln\left(\frac{2 e^{\gamma} \epsilon_c}{\pi T} \right) - \frac{1}{2} \left\lbrace \mathrm{Re} \left[\psi^{(0)}\left(\frac{1}{2} + i \frac{\lambda \| \boldsymbol{g}_{\mathbf{k}} \|}{2 \pi T} \right) \right] - \psi^{(0)}\left(\frac{1}{2} \right) \right\rbrace, \label{SM_GG_01}\\
\frac{1}{\beta} \sum_{\vert \omega_n \vert < \epsilon_c} \int^{\infty}_{- \infty} d\xi G_{-}(- \mathbf{k}, - i\omega_n) G_{-}(\mathbf{k}, i\omega_n) & = \frac{1}{2} \left\lbrace \mathrm{Re} \left[\psi^{(0)}\left(\frac{1}{2} + i \frac{\lambda \| \boldsymbol{g}_{\mathbf{k}} \|}{2 \pi T} \right) \right] - \psi^{(0)}\left(\frac{1}{2} \right) \right\rbrace, \label{SM_GG_02}\\
\frac{1}{\beta} \sum_{\vert \omega_n \vert < \epsilon_c} \int^{\infty}_{- \infty} d\xi G_{\pm}(- \mathbf{k}, - i\omega_n) G_{\mp}(\mathbf{k}, i\omega_n) & = 0, \label{SM_GG_03}
\end{align}
where $\gamma = 0.57721...$ and $\psi^{(0)}(z)$ denote, respectively, the Euler constant and the digamma function. Notice that in light of the result in Eq. \eqref{SM_GG_03}, the gap equations for the singlet and triplet pairing channels become decoupled at linear order.

Finally, if we consider that the gap functions and the pairing interaction are normalized over the Fermi surface as
\begin{gather}
\langle \vert \Delta_s(\mathbf{k}) \vert^2 \rangle_{\mathbf{k}} = \vert \Delta_s \vert^2, \hspace{0.5cm} \langle \| \boldsymbol{\Delta}_t(\mathbf{k}) \|^2 \rangle_{\mathbf{k}} = \vert \Delta_t \vert^2, \\ 
\langle V_{\mathbf{k}, \mathbf{k}'} \Delta_s(\mathbf{k}) \rangle_{\mathbf{k}} = - V_s \Delta_s(\mathbf{k}'), \hspace{0.5cm} \langle V_{\mathbf{k}, \mathbf{k}'} \boldsymbol{\Delta}_t(\mathbf{k}) \rangle_{\mathbf{k}} = - V_t \boldsymbol{\Delta}_t(\mathbf{k}'),
\end{gather}
where $\langle (\cdots) \rangle_{\mathbf{k}} \equiv \int_{\mathcal{S}^2} \frac{d\Omega_{\mathbf{k}}}{4 \pi} (\cdots)$ and $V_s$ ($V_t$) is the pairing coupling for the singlet (triplet) SC state, the gap equations [Eqs. \eqref{SM_GapEq_02} and \eqref{SM_GapEq_03}] become:
\begin{align}
\ln\left(\frac{T_{cs}}{T_{cs,0}}\right) & = - \int_{\mathcal{S}^2} \frac{d\Omega_{\mathbf{k}}}{4\pi} \left\lbrace \mathrm{Re} \left[\psi^{(0)}\left(\frac{1}{2} + i \frac{\lambda \| \boldsymbol{g}_{\mathbf{k}} \|}{2 \pi T_{cs}} \right) \right] - \psi^{(0)}\left(\frac{1}{2} \right) \right\rbrace \vert d_0(\mathbf{k}) \vert^2, \label{SM_GapEq_05} \\
\ln\left(\frac{T_{ct}}{T_{cs,0}}\right) & = - \int_{\mathcal{S}^2} \frac{d\Omega_{\mathbf{k}}}{4\pi} \left\lbrace \mathrm{Re} \left[\psi^{(0)}\left(\frac{1}{2} + i \frac{\lambda \| \boldsymbol{g}_{\mathbf{k}} \|}{2 \pi T_{ct}} \right) \right] - \psi^{(0)}\left(\frac{1}{2} \right) \right\rbrace \vert \boldsymbol{d}(\mathbf{k}) \cdot \widehat{\boldsymbol{g}}_{\mathbf{k}} \vert^2. \label{SM_GapEq_06}
\end{align}
Here, $\pi T_{cs, 0} = 2 e^{\gamma} \epsilon_c e^{-1/(N_F V_s)}$ and $\pi T_{ct, 0} = 2 e^{\gamma} \epsilon_c e^{-1/(N_F V_t)}$ are, respectively, the SC transition temperatures for the singlet and triplet states in the absence of the AM interaction.

To determine $T_{cs}$ in Eq. \eqref{SM_GapEq_05} for a spin-singlet SC phase emerging from an AM state with $B^{-}_{1g}/D_{4h}$ order parameter, we employed the gap functions
\begin{gather}
d_{0, A_{1g}}(\mathbf{k}) = \cos k_x + \cos k_y, \label{SM_SingletGF01} \\
d_{0, B_{1g}}(\mathbf{k}) = \cos k_x - \cos k_y, \label{SM_SingletGF02} \\
d_{0, B_{2g}}(\mathbf{k}) = \sin k_x \sin k_y, \label{SM_SingletGF03}
\end{gather}
which have extended $s$-wave, $d_{x^2 - y^2}$-wave, and $d_{xy}$-wave symmetries, respectively. The subscript $i$ in $d_{0, i}(\mathbf{k})$ refers to the irrep of the $D_{4h}$ point group according to which the $d_{0, i}(\mathbf{k})$ transform. Other choices for $d_{0, i}(\mathbf{k})$ involve cyclic permutations between the $x$, $y$, and $z$ momentum coordinates, because $B^{-}_{1g}/D_{4h}$ AM form factor considered in our numerical calculations is invariant with respect to these operations. On the other hand, the solution for the spin-triplet $T_{ct}$ was obtained for gap functions with $p$-wave symmetry. As explained in the main text, we considered the following possibilities:
\begin{gather}
\boldsymbol{d}_{A_{1u}}(\mathbf{k}) = \sin k_x \hat{\boldsymbol{x}} + \sin k_y \hat{\boldsymbol{y}} + \sin k_z \hat{\boldsymbol{z}}, \label{SM_TripletGF01} \\
\boldsymbol{d}_{A_{2u}}(\mathbf{k}) = - \sin k_y \hat{\boldsymbol{x}} + \sin k_x \hat{\boldsymbol{y}}, \label{SM_TripletGF02} \\
\boldsymbol{d}_{B_{2u}}(\mathbf{k}) = \sin k_y \hat{\boldsymbol{x}} + \sin k_x \hat{\boldsymbol{y}}. \label{SM_TripletGF03}
\end{gather}  
Once again, cyclic permutations of the $x$, $y$, and $z$ momentum coordinates in the expressions for $\boldsymbol{d}_i(\mathbf{k})$ give rise to $p$-wave gap functions with the same value of $T_{ct}$. 

\begin{figure}[t]
\centering
\centering \includegraphics[width=0.8\linewidth,valign=t]{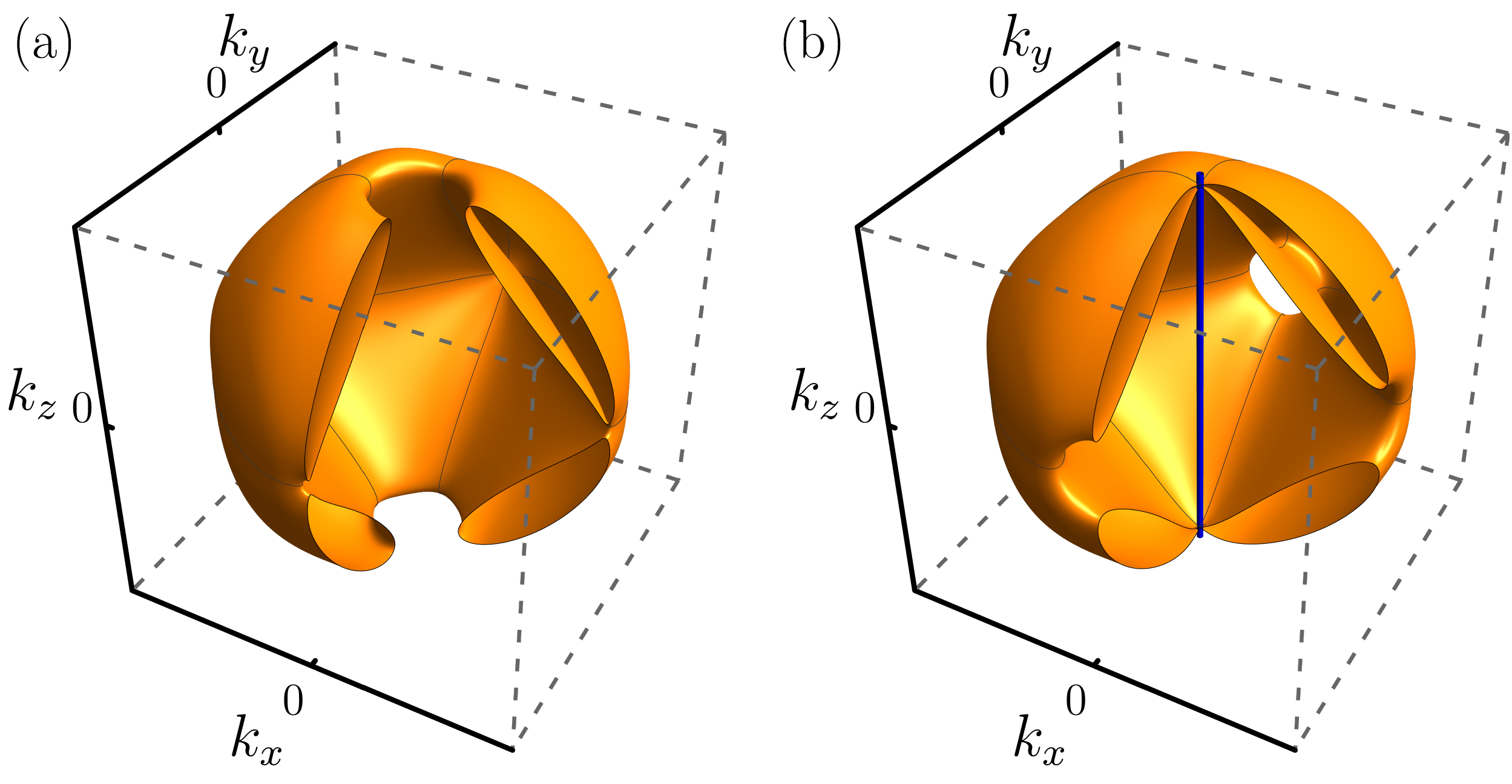} 
\caption{Bogoliubov-Fermi surfaces for a SC state induced by the interplay between spin-singlet pairing and AM interactions. Panels (a) and (b) are obtained with the spin-singlet gap functions $d_{0, A_{1g}}(\mathbf{k}) = \cos k_x + \cos k_y$ (extended $s$-wave symmetry) and $d_{0, B_{1g}}(\mathbf{k}) = \cos k_x - \cos k_y$ ($d_{x^2 - y^2}$-wave symmetry). The vertical blue line in (b) represents the only nodal line of the parent AM metal with $B^{-}_{1g}/D_{4h}$ form factor which is not gapped out in the $d_{x^2 - y^2}$-wave SC state. The intersection between this nodal line with the Bogoliubov-Fermi surface refers to the positions of the Weyl pinching points. In addition, the parts of the Bogoliubov-Fermi surfaces within the domain $\mathcal{D} = \{(k_x, k_y, k_z) \in \mathbb{R}^3 \vert k_x > 0 \land k_y < 0 \}$ were excluded in order to permit  visualization of their internal structure.}\label{BF_Surface}
\end{figure}

\section{Nodal manifolds of the superconducting phase}
\noindent

The electronic dispersions for the SC state in the presence of AM interactions are obtained by the diagonalization of the lattice BdG Hamiltonian: 
\begin{equation}
\widehat{\mathcal{H}}_{\mathrm{BdG}}(\mathbf{k}) = \begin{pmatrix} \widehat{\mathcal{H}}_0(\mathbf{k}) & \widehat{\Delta}(\mathbf{k}) \\ \widehat{\Delta}^{\dagger}(\mathbf{k}) & - \hat{\sigma}_y \widehat{\mathcal{H}}^{T}_0(- \mathbf{k}) \hat{\sigma}_y \end{pmatrix}.
\end{equation}
As explained in the main text, $\widehat{\mathcal{H}}_{\mathrm{BdG}}(\mathbf{k})$ has charge-conjugation symmetry $C$: $\widehat{U}_C \widehat{\mathcal{H}}^{T}_{\mathrm{BdG}}(- \mathbf{k}) \widehat{U}^{\dagger}_C = - \widehat{\mathcal{H}}_{\mathrm{BdG}}(\mathbf{k})$, where $\widehat{U}_C = \hat{\tau}_y \otimes \hat{\sigma}_y$ and $\hat{\tau}_i$ are the Pauli matrices in Nambu space. As a result, the dispersions evaluate to $E_{\nu, \pm}(\mathbf{k}) = \pm E_{\nu}(\mathbf{k})$ with $E_{\nu = \pm}(\mathbf{k})$ representing a positive function. For a pure spin-singlet SC state, $E_{\nu}(\mathbf{k})$ can be calculated exactly, resulting in the expression
\begin{equation}
E_{\nu}(\mathbf{k}) = \vert \sqrt{\xi^2(\mathbf{k}) + \vert \Delta_s(\mathbf{k}) \vert^2} + \nu \lambda \| \boldsymbol{g}_{\mathbf{k}} \| \vert.
\end{equation}
After determining the zeros of $E_{-}(\mathbf{k}) = \vert \sqrt{\xi^2(\mathbf{k}) + \vert \Delta_s(\mathbf{k}) \vert^2} - \lambda \| \boldsymbol{g}_{\mathbf{k}} \| \vert$, we find that the two Fermi-surface sheets defining the normal AM metal do not vanish in the spin-singlet SC state, but instead form a Fermi surface of neutral Bogoliubov quasiparticles known as Bogoliubov-Fermi surfaces (see Fig. \ref{BF_Surface}) \cite{Agterberg-PRL(2017),Brydon-PRB(2018)}. To address the stability of these surfaces, we start by noting that for spin-singlet SC states the BdG Hamiltonian has parity symmetry $P$: $\widehat{U}_P \widehat{\mathcal{H}}_{\mathrm{BdG}}(- \mathbf{k}) \widehat{U}^{\dagger}_P = \widehat{\mathcal{H}}_{\mathrm{BdG}}(\mathbf{k})$ with $\widehat{U}_P = \hat{\tau}_0 \otimes \hat{\sigma}_0$. Consequently, $\widehat{\mathcal{H}}_{\mathrm{BdG}}(\mathbf{k})$ is also invariant under the effect of $CP$ symmetry, since $\widehat{U}_{CP} \widehat{\mathcal{H}}^T_{\mathrm{BdG}}(\mathbf{k}) \widehat{U}^{\dagger}_{CP} = - \widehat{\mathcal{H}}_{\mathrm{BdG}}(\mathbf{k})$ with $\widehat{U}_{CP} = \hat{\tau}_y \otimes \hat{\sigma}_y$. Since $(CP)^2 \equiv (\widehat{U}_{CP} \mathcal{K})^2 = + \mathbb{1}$ and the system breaks time-reversal symmetry, the Bogoliubov-Fermi surfaces acquire a $\mathbb{Z}_2$ topological charge which makes them stable against $CP$-preserving perturbations \cite{Kobayashi-PRB(2014),Zhao-PRL(2016)}. The expression for this topological invariant is obtained following the method devised in Refs. \cite{Agterberg-PRL(2017),Brydon-PRB(2018)}. As $\widehat{U}_{CP} = \hat{\tau}_y \otimes \hat{\sigma}_y$ is symmetric, it can be diagonalized by a unitary congruence $\widehat{U}_{CP} = \widehat{Q} \widehat{\Lambda} \widehat{Q}^T$, where $\widehat{\Lambda}$ is a diagonal matrix and $\widehat{Q}$ is unitary. Among the solutions for these matrices, we choose 
\begin{equation}
\widehat{Q} = \frac{1}{\sqrt{2}}\begin{pmatrix} \hat{\sigma}_z & - \hat{\sigma}_z \\ \hat{\sigma}_x & \hat{\sigma}_x \end{pmatrix} \hspace{0.5cm} \text{and} \hspace{0.5cm} \widehat{\Lambda} = \begin{pmatrix} -\hat{\sigma}_0 & \mathbb{0} \\ \mathbb{0} & \hat{\sigma}_0 \end{pmatrix}.
\end{equation}
After finding the unitary matrix $\sqrt{\widehat{\Lambda}}$ from the equation $\sqrt{\widehat{\Lambda}} \cdot \sqrt{\widehat{\Lambda}} \equiv \widehat{\Lambda}$, one can prove that the unitary transformation of the BdG Hamiltonian with respect to $\widehat{\Omega} \equiv \sqrt{\widehat{\Lambda}}^{\dagger} \widehat{Q}^{\dagger}$ yields:
\begin{equation}
\left[ \widehat{\Omega} \widehat{\mathcal{H}}_{\mathrm{BdG}}(\mathbf{k}) \widehat{\Omega}^{\dagger} \right]^T = - \widehat{\Omega} \widehat{\mathcal{H}}_{\mathrm{BdG}}(\mathbf{k}) \widehat{\Omega}^{\dagger}.
\end{equation}
Because $\widehat{\Omega} \widehat{\mathcal{H}}_{\mathrm{BdG}}(\mathbf{k}) \widehat{\Omega}^{\dagger}$ is anti-symmetric, one can define the Pfaffian $P(\mathbf{k}) \equiv \mathrm{Pf}[\widehat{\Omega} \widehat{\mathcal{H}}_{\mathrm{BdG}}(\mathbf{k}) \widehat{\Omega}^{\dagger}]$. Then we obtain:
\begin{equation}
P(\mathbf{k}) = \xi^2_{\mathbf{k}} + \vert \Delta_s(\mathbf{k}) \vert^2 - (\lambda \| \boldsymbol{g}_\mathbf{k} \|)^2.
\end{equation}
As already explained in the main text, $P(\mathbf{k})$ changes sign at $E_{-}(\mathbf{k}) = 0$ when the AM interaction is finite. Consequently, the $\mathbb{Z}_2$ topological invariant associated with the Bogoliubov-Fermi surface turns out to be $(-1)^l = \mathrm{sgn}[P(\mathbf{k}_{+}) P(\mathbf{k}_{-})]$, with $\mathbf{k}_{-}$ and $\mathbf{k}_{+}$ representing points inside and outside of the Bogoliubov-Fermi surface, respectively.

\begin{figure}[t]
\centering
\centering \includegraphics[width=0.85\linewidth,valign=t]{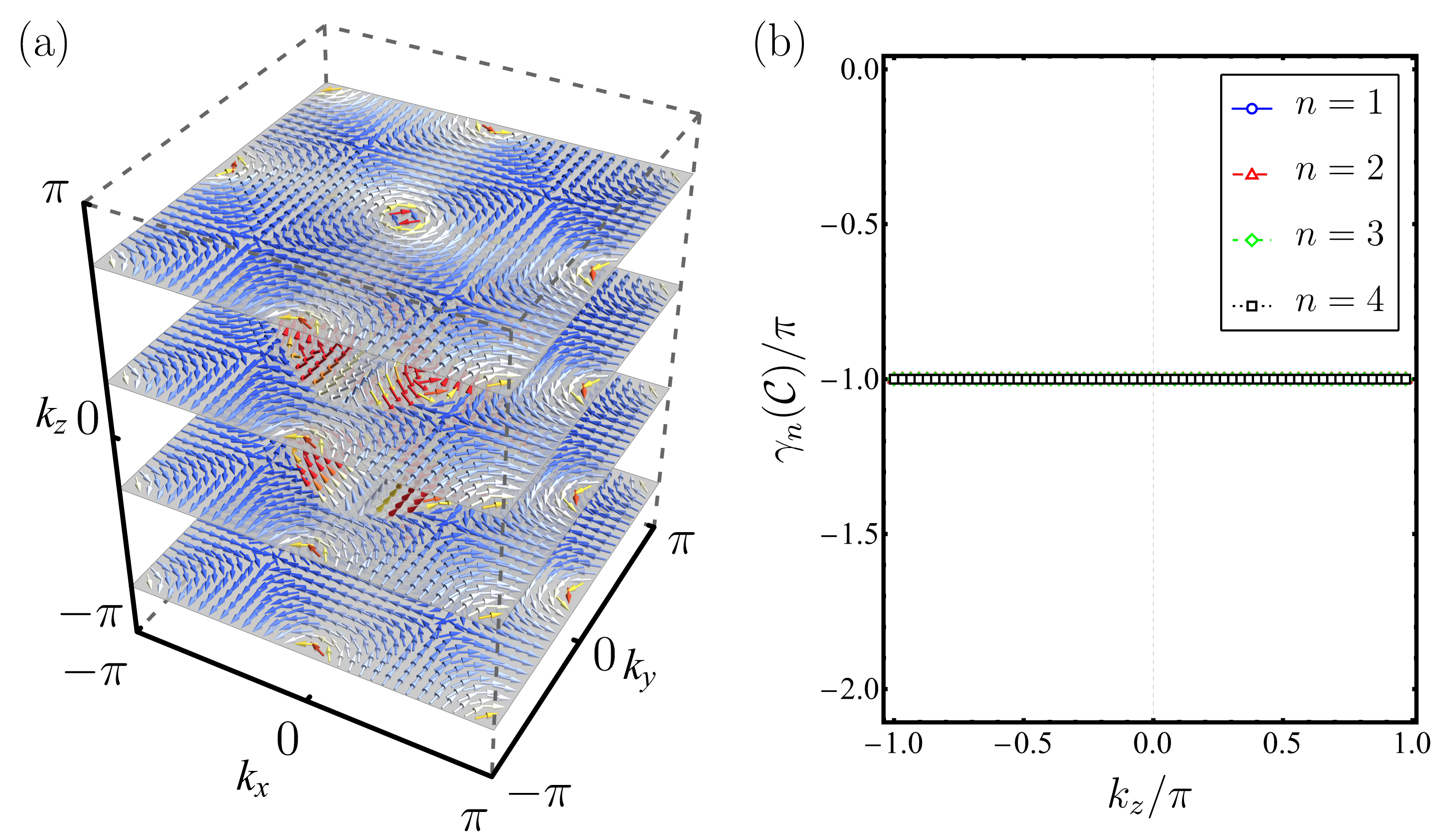}
\caption{(a) Dependence of the Berry connection $\mathbf{A}_n(\mathbf{k}) = i\langle \psi_n(\mathbf{k}) \vert \nabla_{\mathbf{k}} \vert \psi_n(\mathbf{k}) \rangle$ on some $k_z$-fixed planes for the lowest-energy eigenstate $\vert \psi_n(\mathbf{k}) \rangle$ of the BdG Hamiltonian in a pure spin-triplet SC phase described by the $\boldsymbol{d}_{B_{2u}}(\mathbf{k})$ gap function with $f$-wave symmetry. Notice that the nodal line $k_x = k_y$ of the energy dispersions $E_{\nu, \pm}(\mathbf{k})$ represents a vortex of the Berry connection. (b) This feature becomes clearer with the evaluation of the Berry phase around an infinitesimal circumference $\mathcal{C}$ perpendicular to the nodal line $k_x = k_y = 0$, which yields $\gamma_{n}(\mathcal{C}) = - \pi$.}\label{Berry_Phase}
\end{figure}

In contrast, when the SC phase is described by a spin-triplet order parameter, the energy dispersions $E_{\nu}(\mathbf{k})$ become:
\begin{equation}
E_{\nu}(\mathbf{k}) = 
\sqrt{\xi^2_{\mathbf{k}} + \| \boldsymbol{\Delta}_t(\mathbf{k}) \|^2 + (\lambda \| \boldsymbol{g}_{\mathbf{k}} \|)^2 + 2 \nu \lambda \sqrt{\xi^2_{\mathbf{k}} \| \boldsymbol{g}_{\mathbf{k}} \|^2 + | \boldsymbol{\Delta}_t(\mathbf{k}) \cdot \boldsymbol{g}_{\mathbf{k}}|^2}}.
\end{equation}
Next, we consider spin-triplet states with gap functions $\boldsymbol{d}(\mathbf{k}) = \boldsymbol{u}(\mathbf{k}) \times \boldsymbol{g}_{\mathbf{k}}$, which, as emphasized in the main text, yields $T_{ct} = T_{ct, 0}$ regardless of the intensity of $\lambda$. Under this assumption, the dispersions evaluate to:
\begin{equation}
E_{\pm}(\mathbf{k}) = \sqrt{(\vert \xi_{\mathbf{k}}\vert \pm \lambda \| \boldsymbol{g}_{\mathbf{k}} \|)^2 + \| \Delta_t \boldsymbol{d}(\mathbf{k}) \|^2}.
\end{equation}
Thus the zero-energy manifold of this spin-triplet state occurs at the intersection of the nodal lines $\boldsymbol{g}_{\mathbf{k}} = \boldsymbol{0}$ and $\boldsymbol{u}(\mathbf{k}) \times \boldsymbol{g}_{\mathbf{k}} = 0$ with the normal AM Fermi surface $(\vert \xi_{\mathbf{k}}\vert - \lambda \|\boldsymbol{g}_{\mathbf{k}} \|) = 0$. Since the spin-triplet SC phase breaks the parity symmetry, the gapless excitations of this state are entirely described by Weyl pinching points.

To demonstrate that these nodal lines have non-trivial topology, we proceed with the computation of their Berry phase. In this specific case, we use the $f$-wave gap function $\boldsymbol{d}_{B_{2u}}(\mathbf{k})$ obtained in the main text. The behavior of the Berry connection $\mathbf{A}_n(\mathbf{k}) = i\langle \psi_n(\mathbf{k}) \vert \nabla_{\mathbf{k}} \vert \psi_n(\mathbf{k}) \rangle$ for the eigenstate $\vert \psi_n(\mathbf{k}) \rangle$ related to the lowest-energy dispersion is illustrated in Fig. \ref{Berry_Phase}(a). Clearly, the nodal lines of the BdG Hamiltonian behave as vortices of the Berry connection. To quantify this statement, we compute the Berry phase $\gamma_n(\mathcal{C}) = \oint_{\mathcal{C}} d\mathbf{k} \cdot \mathbf{A}_n(\mathbf{k})$ for a circular path $\mathcal{C}$ around the nodal line $k_x = k_y = 0$. As displayed in Fig. \ref{Berry_Phase}(b), we obtain $\gamma_n(\mathcal{C}) = - \pi$ for $k_z \in [ -\pi, \pi [$. Since the Brillouin zone contains vortex-like and antivortex-like nodal lines, the general result for the Berry phase is $\gamma_n(\mathcal{C}) = \pm \pi$. Due to the non-trivial value of $\gamma_n(\mathcal{C})$, the nodal lines in this spin-triplet SC phase turn out to be topological stable.

\section{Landau-Ginzburg free energy of the superconducting phase}
\noindent

To determine the Ginzburg-Landau expansion of the effective action $\mathcal{S}_{\mathrm{eff}}[\widehat{\Delta}^{\dagger}, \widehat{\Delta}]$ \cite{Altland-CUP(2010)}, we first rewrite the interacting propagator as $\widehat{\mathcal{G}}^{-1} = \widehat{\mathcal{G}}^{-1}_0 + \widehat{\varDelta}$, which then leads to:
\begin{equation}\label{SM_TraceExp}
\mathrm{Tr}\ln\left(\widehat{\mathcal{G}}^{-1}\right) = \mathrm{Tr}\ln\left(\widehat{\mathcal{G}}^{-1}_0\right) - \sum^{\infty}_{\ell = 1}\frac{1}{2\ell}\mathrm{Tr}\left(\widehat{\mathcal{G}}_0 \widehat{\varDelta}\right)^{2\ell},
\end{equation}
where
\begin{equation}
\widehat{\mathcal{G}}^{-1}_0 \equiv \begin{pmatrix} \widehat{G}^{-1}_{0, p} & \mathbb{0} \\ \mathbb{0} & \widehat{G}^{-1}_{0, h} \end{pmatrix}, \hspace{0.5cm} \widehat{\varDelta} \equiv \begin{pmatrix} \mathbb{0} & \widehat{\Delta} \\ \widehat{\Delta}^{\dagger} & \mathbb{0} \end{pmatrix}.
\end{equation}

The first trace defined on the r.h.s. of Eq. \eqref{SM_TraceExp} is given by
\begin{align}
\mathrm{Tr}\left(\widehat{\mathcal{G}}_0 \widehat{\varDelta}\right)^2 & = \frac{1}{2} \sum_{n, \mathbf{k}}\mathrm{tr}
\begin{pmatrix}
\widehat{G}_{0, p}(\mathbf{k}, i\omega_n) \widehat{\Delta}(\mathbf{k})\widehat{G}_{0, h}(\mathbf{k}, i\omega_n)\widehat{\Delta}^{\dagger}(\mathbf{k}) & \mathbb{0} \\
\mathbb{0} & \widehat{G}_{0, h}(\mathbf{k}, i\omega_n) \widehat{\Delta}^{\dagger}(\mathbf{k})\widehat{G}_{0, p}(\mathbf{k}, i\omega_n)\widehat{\Delta}(\mathbf{k})
\end{pmatrix} \nonumber\\
& = \sum_{n, \mathbf{k}}\mathrm{tr}[\widehat{G}_{0, p}(\mathbf{k}, i\omega_n) \widehat{\Delta}(\mathbf{k})\widehat{G}_{0, h}(\mathbf{k}, i\omega_n)\widehat{\Delta}^{\dagger}(\mathbf{k})].
\end{align}
Using the substitutions in Eqs. \eqref{SM_PartProp} and \eqref{SM_HoleProp}, the expression for $\mathrm{Tr}\left(\widehat{\mathcal{G}}_0 \widehat{\varDelta}\right)^2$ simplifies to:
\begin{align}
\mathrm{Tr}\left(\widehat{\mathcal{G}}_0 \widehat{\varDelta}\right)^2 & = - 2 N_F\beta \int_{\mathcal{S}^2} \frac{d\Omega_{\mathbf{k}}}{4\pi} \bigg\lbrace \frac{1}{\beta}\sum_{\vert \omega_n \vert < \epsilon_c} \int^{\infty}_{-\infty} d\xi(\widetilde{G}_{+} G_{+} - \widetilde{G}_{-} G_{-}) \vert \Delta_s(\mathbf{k})\vert^2 + \frac{1}{\beta}\sum_{\vert \omega_n \vert < \epsilon_c} \int^{\infty}_{-\infty} d\xi \widetilde{G}_{+} G_{+} \| \boldsymbol{\Delta}_t(\mathbf{k}) \|^2 \nonumber \\
& + \frac{1}{\beta}\sum_{\vert \omega_n \vert < \epsilon_c} \int^{\infty}_{-\infty} d\xi \widetilde{G}_{-} G_{-}[\| \widehat{\boldsymbol{g}}_{\mathbf{k}} \times \boldsymbol{\Delta}_t(\mathbf{k}) \|^2 - \vert \widehat{\boldsymbol{g}}_{\mathbf{k}} \cdot \boldsymbol{\Delta}_t(\mathbf{k}) \vert^2] \bigg\rbrace.
\end{align}
Because $\| \widehat{\boldsymbol{g}}_{\mathbf{k}} \times \boldsymbol{\Delta}_t(\mathbf{k}) \|^2 = \| \boldsymbol{\Delta}_t(\mathbf{k}) \|^2 - \vert \widehat{\boldsymbol{g}}_{\mathbf{k}} \cdot \boldsymbol{\Delta}_t(\mathbf{k}) \vert^2$ and using the results derived in the previous section, it becomes straightforward to express $\mathrm{Tr}\left(\widehat{\mathcal{G}}_0 \widehat{\varDelta}\right)^2$ as
\begin{equation}
\mathrm{Tr}\left(\widehat{\mathcal{G}}_0 \widehat{\varDelta}\right)^2 = 2 \beta (\chi_1 \vert \Delta_s\vert^2 + \chi_2 \vert \Delta_t \vert^2),
\end{equation}
where
\begin{align}
\chi_1 & = N_F \int_{\mathcal{S}^2} \frac{d \Omega_{\mathbf{k}}}{4 \pi} \bigg\lbrace \mathrm{Re}\bigg[\psi^{(0)}\bigg(\frac{1}{2} + i \frac{\lambda \| \boldsymbol{g}_{\mathbf{k}} \|}{2 \pi T} \bigg)\bigg] - \psi^{(0)}\bigg(\frac{1}{2}\bigg) - \ln\bigg(\frac{2 e^{\gamma} \epsilon_c}{\pi T} \bigg) \bigg\rbrace d^2_0(\mathbf{k}), \\
\chi_2 & = N_F \int_{\mathcal{S}^2} \frac{d \Omega_{\mathbf{k}}}{4 \pi} \bigg\lbrace \bigg\lbrace \mathrm{Re}\bigg[\psi^{(0)}\bigg(\frac{1}{2} + i \frac{\lambda \| \boldsymbol{g}_{\mathbf{k}} \|}{2 \pi T} \bigg)\bigg] - \psi^{(0)}\bigg(\frac{1}{2}\bigg) \bigg\rbrace [\widehat{\boldsymbol{g}}_{\mathbf{k}}\cdot \boldsymbol{d}(\mathbf{k})]^2 - \ln\bigg(\frac{2 e^{\gamma} \epsilon_c}{\pi T} \bigg) \boldsymbol{d}^2(\mathbf{k}) \bigg\rbrace.
\end{align}

The expression for the second trace on the r.h.s. of Eq. \eqref{SM_TraceExp} reads:
\begin{align}
\mathrm{Tr}\left(\widehat{\mathcal{G}}_0 \widehat{\varDelta}\right)^4 & = \frac{1}{2} \sum_{n, \mathbf{k}}\mathrm{tr}
\begin{pmatrix}
[\widehat{G}_{0, p}(\mathbf{k}, i\omega_n) \widehat{\Delta}(\mathbf{k})\widehat{G}_{0, h}(\mathbf{k}, i\omega_n)\widehat{\Delta}^{\dagger}(\mathbf{k})]^2 & \mathbb{0} \\
\mathbb{0} & [\widehat{G}_{0, h}(\mathbf{k}, i\omega_n) \widehat{\Delta}^{\dagger}(\mathbf{k})\widehat{G}_{0, p}(\mathbf{k}, i\omega_n)\widehat{\Delta}(\mathbf{k})]^2
\end{pmatrix} \nonumber\\
& = \sum_{n, \mathbf{k}}\mathrm{tr} \lbrace [\widehat{G}_{0, p}(\mathbf{k}, i\omega_n) \widehat{\Delta}(\mathbf{k})\widehat{G}_{0, h}(\mathbf{k}, i\omega_n)\widehat{\Delta}^{\dagger}(\mathbf{k})]^2 \rbrace.
\end{align}
Henceforth, we will take into account triplet SC states satisfying the constraint 
\begin{equation}
\boldsymbol{\Delta}^{*}_t(\mathbf{k}) \cdot [\widehat{\boldsymbol{g}}_{\mathbf{k}} \times \boldsymbol{\Delta}_t(\mathbf{k})] = 0.
\end{equation}
As a result, we are able to find:
\begin{align}
\mathrm{Tr}\left(\widehat{\mathcal{G}}_0 \widehat{\varDelta}\right)^4 & = N_F \beta \int_{\mathcal{S}^2} \frac{d \Omega_{\mathbf{k}}}{4 \pi} \lbrace \eta_1(\mathbf{k})\vert \Delta_s(\mathbf{k})\vert^4 + \eta_2(\mathbf{k})\vert \boldsymbol{\Delta}^2_t(\mathbf{k})\vert^2 + \eta_3(\mathbf{k})\vert \widehat{\boldsymbol{g}}_{\mathbf{k}} \cdot \boldsymbol{\Delta}_t(\mathbf{k})\vert^4 + \eta_4(\mathbf{k}) \vert \boldsymbol{\Delta}^2_t(\mathbf{k})\vert \vert \widehat{\boldsymbol{g}}_{\mathbf{k}} \cdot \boldsymbol{\Delta}_t(\mathbf{k})\vert^2 \nonumber \\
& + \zeta_1(\mathbf{k}) \vert \Delta_s(\mathbf{k})\vert^2 \vert \boldsymbol{\Delta}^2_t(\mathbf{k})\vert + \zeta_2(\mathbf{k}) \vert \Delta_s(\mathbf{k})\vert^2 \vert \widehat{\boldsymbol{g}}_{\mathbf{k}} \cdot \boldsymbol{\Delta}_t(\mathbf{k})\vert^2 + \rho_1(\mathbf{k})\left( \Delta^2_s(\mathbf{k}) [\widehat{\boldsymbol{g}}^{*}_{\mathbf{k}} \cdot \boldsymbol{\Delta}^{*}_t(\mathbf{k})]^2 + [\Delta^{*}_s(\mathbf{k})]^2 [\widehat{\boldsymbol{g}}_{\mathbf{k}} \cdot \boldsymbol{\Delta}_t(\mathbf{k})]^2 \right) \nonumber \\
& + \rho_2(\mathbf{k})\left( \Delta^2_s(\mathbf{k}) [\boldsymbol{\Delta}^2_t(\mathbf{k})]^{*} + [\Delta^{*}_s(\mathbf{k})]^2 \boldsymbol{\Delta}^2_t(\mathbf{k}) \right) \rbrace,
\end{align}
where
\begin{gather}
\eta_1(\mathbf{k}) \equiv \frac{1}{\beta} \sum^{\infty}_{n = - \infty} \frac{\pi \vert \omega_n \vert [\omega^2_n - 3(\lambda \| \boldsymbol{g}_{\mathbf{k}} \|)^2]}{[(\lambda \| \boldsymbol{g}_{\mathbf{k}} \|)^2 + \omega^2_n]^3}, \\
\eta_2(\mathbf{k}) \equiv \frac{1}{\beta} \sum^{\infty}_{n = - \infty} \frac{\pi}{\vert \omega_n \vert^3}, \\
\eta_3(\mathbf{k}) \equiv \frac{1}{\beta} \sum^{\infty}_{n = - \infty} \frac{\pi (\lambda \| \boldsymbol{g}_{\mathbf{k}} \|)^4 [(\lambda \| \boldsymbol{g}_{\mathbf{k}} \|)^2 + 5\omega^2_n]}{\vert \omega_n \vert^3 [(\lambda \| \boldsymbol{g}_{\mathbf{k}} \|)^2 + \omega^2_n]^3}, \\
\eta_4(\mathbf{k}) \equiv -\frac{1}{\beta}\sum^{\infty}_{n = - \infty} \frac{2\pi (\lambda \| \boldsymbol{g}_{\mathbf{k}} \|)^2[(\lambda \| \boldsymbol{g}_{\mathbf{k}} \|)^4 + 4\omega^2_n (\lambda \| \boldsymbol{g}_{\mathbf{k}} \|)^2 + 3\omega^4_n]}{\vert\omega_n\vert^3 [(\lambda \| \boldsymbol{g}_{\mathbf{k}} \|)^2 + \omega^2_n]^3}, \\
\zeta_1(\mathbf{k}) \equiv \frac{1}{\beta}\sum^{\infty}_{n = - \infty} \frac{4 \pi \vert\omega_n\vert}{[(\lambda \| \boldsymbol{g}_{\mathbf{k}} \|)^2 + \omega^2_n]^2}, \\
\zeta_2(\mathbf{k}) \equiv - \frac{1}{\beta}\sum^{\infty}_{n = - \infty} \frac{16 \pi (\lambda \| \boldsymbol{g}_{\mathbf{k}} \|)^2 \vert\omega_n\vert}{[(\lambda \| \boldsymbol{g}_{\mathbf{k}} \|)^2 + \omega^2_n]^2}, \\
\rho_1(\mathbf{k}) \equiv - \frac{1}{\beta}\sum^{\infty}_{n = - \infty} \frac{\pi (\lambda \| \boldsymbol{g}_{\mathbf{k}} \|)^2 [(\lambda \| \boldsymbol{g}_{\mathbf{k}} \|)^2 + 5 \omega^2_n]}{\vert\omega_n\vert [(\lambda \| \boldsymbol{g}_{\mathbf{k}} \|)^2 + \omega^2_n]^3}, \\
\rho_2(\mathbf{k}) \equiv \frac{1}{\beta}\sum^{\infty}_{n = - \infty} \frac{\pi}{\vert\omega_n\vert [(\lambda \| \boldsymbol{g}_{\mathbf{k}} \|)^2 + \omega^2_n]}.
\end{gather}
After the substitutions $\Delta_s(\mathbf{k}) = \Delta_s d_0(\mathbf{k})$ and $\boldsymbol{\Delta}_t = \Delta_t \boldsymbol{d}(\mathbf{k})$, the equation for $\mathrm{Tr}\left(\widehat{\mathcal{G}}_0 \widehat{\varDelta}\right)^4$ simplifies to:
\begin{equation}
\mathrm{Tr}\left(\widehat{\mathcal{G}}_0 \widehat{\varDelta}\right)^4 = 4\beta \left\lbrace \beta_1 \vert\Delta_s\vert^4 + \beta_2 \vert\Delta_t\vert^4 + 2\gamma_1 \vert\Delta_s\vert^2 \vert\Delta_t\vert^2 + \gamma_2[\Delta^2_s(\Delta^{*}_t)^2 + (\Delta^{*}_s)^2 \Delta^2_t] \right\rbrace,
\end{equation}
where
\begin{align}
\beta_1 & \equiv \frac{N_F}{4} \int_{\mathcal{S}^2} \frac{d\Omega_{\mathbf{k}}}{4\pi} d^4_0(\mathbf{k}) \eta_1(\mathbf{k}) \nonumber \\
& = - \frac{N_F}{32 \pi^2 T^2} \int_{\mathcal{S}^2} \frac{d\Omega_{\mathbf{k}}}{4 \pi} d^4_0(\mathbf{k}) \mathrm{Re}\bigg[\psi^{(2)}\bigg(\frac{1}{2} + i \frac{\lambda \| \boldsymbol{g}_{\mathbf{k}} \|}{2 \pi T} \bigg) \bigg], \\
\beta_2 & \equiv \frac{N_F}{4} \int_{\mathcal{S}^2} \frac{d\Omega_{\mathbf{k}}}{4\pi} \lbrace [\boldsymbol{d}^2(\mathbf{k})]^2 \eta_2(\mathbf{k}) + [\widehat{\boldsymbol{g}}_{\mathbf{k}} \cdot \boldsymbol{d}(\mathbf{k})]^4\eta_3(\mathbf{k}) + \boldsymbol{d}^2(\mathbf{k})[\widehat{\boldsymbol{g}}_{\mathbf{k}} \cdot \boldsymbol{d}(\mathbf{k})]^2\eta_4(\mathbf{k}) \rbrace \nonumber \\
& = \frac{7 \zeta(3) N_F}{16 \pi^2 T^2} \int_{\mathcal{S}^2} \frac{d\Omega_{\mathbf{k}}}{4 \pi} [\boldsymbol{d}^2(\mathbf{k})]^2 - \frac{N_F}{32 \pi^2 T^2} \int_{\mathcal{S}^2} \frac{d\Omega_{\mathbf{k}}}{4 \pi} \frac{[\widehat{\boldsymbol{g}}_{\mathbf{k}} \cdot \boldsymbol{d}(\mathbf{k})]^4}{\| \boldsymbol{g}_\mathbf{k} \|^2} \bigg\lbrace  \| \boldsymbol{g}_\mathbf{k} \|^2 \mathrm{Re}\bigg[\psi^{(2)}\bigg(\frac{1}{2} + i \frac{\lambda \| \boldsymbol{g}_{\mathbf{k}} \|}{2 \pi T} \bigg) \bigg] + \psi^{(2)}\bigg(\frac{1}{2} \bigg) \| \boldsymbol{g}_\mathbf{k} \|^2 \nonumber \\
& - \frac{16 \pi^2 T^2}{\lambda^2} \mathrm{Re}\bigg[\psi^{(0)}\bigg(\frac{1}{2} + i \frac{\lambda \| \boldsymbol{g}_{\mathbf{k}} \|}{2 \pi T} \bigg) \bigg] - \frac{8 \pi \| \boldsymbol{g}_{\mathbf{k}} \| T}{\lambda} \mathrm{Im}\bigg[\psi^{(1)}\bigg(\frac{1}{2} + i \frac{\lambda \| \boldsymbol{g}_{\mathbf{k}} \|}{2 \pi T} \bigg) \bigg] + 16 \pi^2 \frac{T^2}{\lambda^2} \psi^{(0)}\bigg(\frac{1}{2} \bigg) \bigg\rbrace \nonumber \\
& + \frac{N_F}{8 \pi^2 T^2} \int_{\mathcal{S}^2} \frac{d\Omega_{\mathbf{k}}}{4 \pi} \frac{\| \boldsymbol{d}(\mathbf{k}) \|^2 [\widehat{\boldsymbol{g}}_{\mathbf{k}} \cdot \boldsymbol{d}(\mathbf{k})]^2}{\| \boldsymbol{g}_{\mathbf{k}} \|^2} \bigg\lbrace 4 \pi^2 \psi^{(0)}\bigg(\frac{1}{2} \bigg) \frac{T^2}{\lambda^2} - 7 \zeta(3) \| \boldsymbol{g}_{\mathbf{k}} \|^2 - 4 \pi^2 \frac{T^2}{\lambda^2} \mathrm{Re}\bigg[\psi^{(0)}\bigg(\frac{1}{2} + i \frac{\lambda \| \boldsymbol{g}_{\mathbf{k}} \|}{2 \pi T} \bigg) \bigg] \nonumber \\
& - 2 \pi \| \boldsymbol{g}_ {\mathbf{k}} \| \frac{T}{\lambda} \mathrm{Im}\bigg[\psi^{(1)}\bigg(\frac{1}{2} + i \frac{\lambda \| \boldsymbol{g}_{\mathbf{k}} \|}{2 \pi T} \bigg) \bigg] \bigg\rbrace, 
\end{align}
\begin{align}
\gamma_1 & \equiv \frac{N_F}{8} \int_{\mathcal{S}^2} \frac{d\Omega_{\mathbf{k}}}{4\pi} d^2_0(\mathbf{k})\lbrace \boldsymbol{d}^2(\mathbf{k}) \zeta_1(\mathbf{k}) + [\widehat{\boldsymbol{g}}_{\mathbf{k}} \cdot \boldsymbol{d}(\mathbf{k})]^2 \zeta_2(\mathbf{k}) \rbrace \nonumber \\
& = - \frac{N_F}{8} \int_{\mathcal{S}^2} \frac{d\Omega_{\mathbf{k}}}{4 \pi} d^2_0(\mathbf{k}) \bigg\lbrace \bigg(\frac{\boldsymbol{d}^2(\mathbf{k}) - [\widehat{\boldsymbol{g}}_{\mathbf{k}} \cdot \boldsymbol{d}(\mathbf{k})]^2}{\pi \lambda T \| \boldsymbol{g}_{\mathbf{k}} \|} \bigg) \mathrm{Im}\bigg[\psi^{(1)}\bigg(\frac{1}{2} + i \frac{\lambda \| \boldsymbol{g}_{\mathbf{k}} \|}{2 \pi T} \bigg) \bigg] + \frac{[\widehat{\boldsymbol{g}}_{\mathbf{k}} \cdot \boldsymbol{d}(\mathbf{k})]^2}{2 \pi^2 T^2}  \mathrm{Re}\bigg[\psi^{(2)}\bigg(\frac{1}{2} + i \frac{\lambda \| \boldsymbol{g}_{\mathbf{k}} \|}{2 \pi T} \bigg) \bigg] \bigg\rbrace, \\
\gamma_2 & \equiv \frac{N_F}{4} \int_{\mathcal{S}^2} \frac{d\Omega_{\mathbf{k}}}{4\pi} d^2_0(\mathbf{k})\lbrace \boldsymbol{d}^2(\mathbf{k}) \rho_1(\mathbf{k}) + [\widehat{\boldsymbol{g}}_{\mathbf{k}} \cdot \boldsymbol{d}(\mathbf{k})]^2 \rho_2(\mathbf{k}) \rbrace \nonumber \\
& = \frac{N_F}{4} \int_{\mathcal{S}^2} \frac{d\Omega_{\mathbf{k}}}{4 \pi} d^2_0(\mathbf{k}) \bigg\lbrace \bigg(\frac{\boldsymbol{d}^2(\mathbf{k}) - [\widehat{\boldsymbol{g}}_{\mathbf{k}} \cdot \boldsymbol{d}(\mathbf{k})]^2}{\lambda^2 \| \boldsymbol{g}_{\mathbf{k}} \|^2} \bigg) \bigg\lbrace \mathrm{Re}\bigg[\psi^{(0)}\bigg(\frac{1}{2} + i \frac{\lambda \| \boldsymbol{g}_{\mathbf{k}} \|}{2 \pi T} \bigg) \bigg] - \psi^{(0)}\bigg(\frac{1}{2} \bigg) \bigg\rbrace - \frac{[\widehat{\boldsymbol{g}}_{\mathbf{k}} \cdot \boldsymbol{d}(\mathbf{k})]^2}{8 \pi^2 T^2} \nonumber\\
& \times \mathrm{Re}\bigg[\psi^{(2)}\bigg(\frac{1}{2} + i \frac{\lambda \| \boldsymbol{g}_{\mathbf{k}} \|}{2 \pi T} \bigg) \bigg] \bigg\rbrace.
\end{align}
In these equations, $\zeta(3) = 1.20205...$ denotes the Ap\'ery's constant.

\begin{figure}[t]
\centering
\centering \includegraphics[width=1.0\linewidth,valign=t]{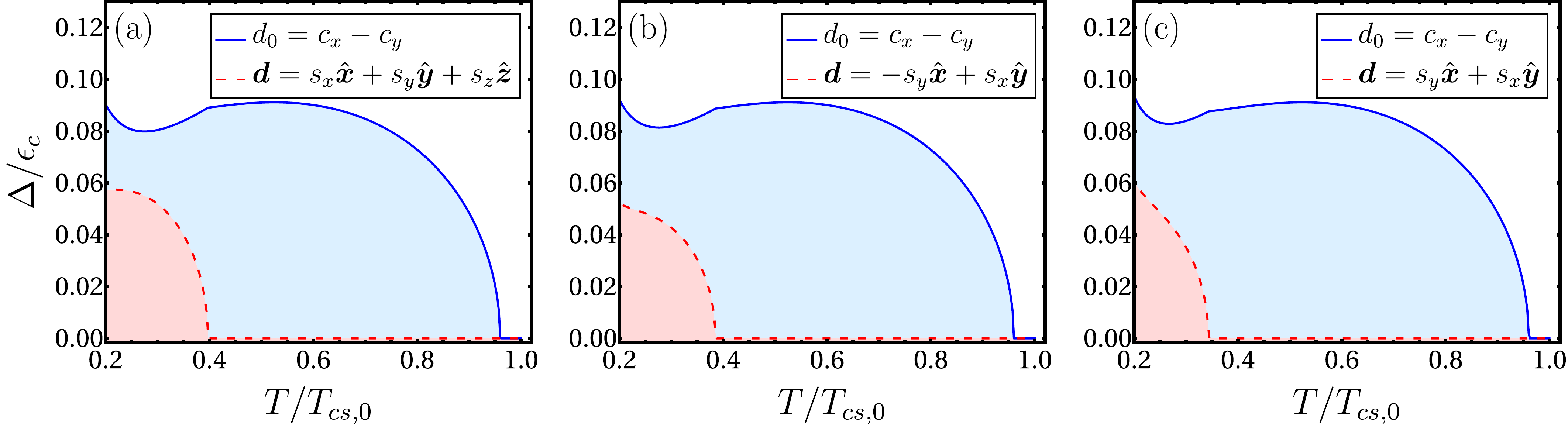} 
\caption{Behavior of the gap amplitudes $\vert \Delta_s \vert$ and $\vert \Delta_t \vert$ as a function of the temperature obtained by the minimization of the Ginzburg-Landau free energy $\mathcal{F}[\Delta_s, \Delta_t]$ in Eq. \eqref{SM_FreeEner02}. Notice that $c_a \equiv \cos k_a$ and $s_a \equiv \sin k_a$ with $a \in \{ x, y, z\}$. We found that the $d_{x^2 - y^2}$-wave SC state coexists with the (a) $A_{1u}$, (b) $B_{1u}$, and (c) $B_{2u}$ $p$-wave states for a finite range of AM interactions. Besides, the phase difference between the $d_{x^2 - y^2}$-wave and $p$-wave gap functions assumes the non-trivial value $\theta = \pm \pi/2$, corresponding to the breaking of time-reversal symmetry.}\label{Gap_Amplitudes}
\end{figure}

According to expressions for $\mathrm{Tr}\left(\widehat{\mathcal{G}}_0 \widehat{\varDelta}\right)^2$ and $\mathrm{Tr}\left(\widehat{\mathcal{G}}_0 \widehat{\varDelta}\right)^4$, the Ginzburg-Landau free energy $\mathcal{F}[\Delta_s, \Delta_t] \equiv \mathcal{S}_{\mathrm{eff}}[\Delta_s, \Delta_t]/\beta$ up to fourth order in both $\Delta_s$ and $\Delta_t$ turns out to be:
\begin{equation}\label{SM_FreeEner01}
\mathcal{F}[\Delta_s, \Delta_t] = \mathcal{F}_0 + \alpha_1 \vert \Delta_s \vert^2 + \alpha_2 \vert \Delta_t \vert^2 + \beta_1 \vert \Delta_s \vert^4 +\beta_2 \vert \Delta_t \vert^4 + 2 \gamma_1 \vert \Delta_s \vert^2 \vert \Delta_t \vert^2 + \gamma_2 [ \Delta^2_s (\Delta^{*}_t)^2 + (\Delta^{*}_s)^2 \Delta^2_t], 
\end{equation}
where 
\begin{align}
\mathcal{F}_0 & = - \frac{1}{\beta} \operatorname{Tr}\ln\left(\widehat{\mathcal{G}}^{-1}_0 \right), \\
\alpha_1 & = N_F \int_{\mathcal{S}^2} \frac{d\Omega_{\mathbf{k}}}{4 \pi} \bigg\lbrace \mathrm{Re}\bigg[ \psi^{(0)}\bigg( \frac{1}{2} + i \frac{\lambda \| \boldsymbol{g}_{\mathbf{k}} \|}{2 \pi T} \bigg) \bigg] - \psi^{(0)} \bigg(\frac{1}{2} \bigg) - \ln \bigg( \frac{2 e^{\gamma} \epsilon_c}{\pi T} \bigg) \bigg\rbrace d^2_0(\mathbf{k}) - \sum_{\mathbf{k}, \mathbf{k}'} d_0(\mathbf{k}') V^{-1}(\mathbf{k}', \mathbf{k}) d_0(\mathbf{k}), \\
\alpha_ 2 & = N_F \int_{\mathcal{S}^2} \frac{d\Omega_{\mathbf{k}}}{4 \pi} \bigg\lbrace \bigg\lbrace \mathrm{Re}\bigg[ \psi^{(0)}\bigg( \frac{1}{2} + i \frac{\lambda \| \boldsymbol{g}_{\mathbf{k}} \|}{2 \pi T} \bigg) \bigg] - \psi^{(0)} \bigg(\frac{1}{2} \bigg) \bigg\rbrace [\widehat{\boldsymbol{g}}_{\mathbf{k}} \cdot \boldsymbol{d}(\mathbf{k})]^2 - \ln \bigg( \frac{2 e^{\gamma} \epsilon_c}{\pi T} \bigg) \boldsymbol{d}^2(\mathbf{k}) \bigg\rbrace \nonumber \\
& - \sum_{\mathbf{k}, \mathbf{k}'}[ d_x(\mathbf{k}') V^{-1}_{\mathbf{k}', \mathbf{k}} d_x(\mathbf{k}) + d_y(\mathbf{k}') V^{-1}_{\mathbf{k}', \mathbf{k}} d_y(\mathbf{k}) + d_z(\mathbf{k}') V^{-1}_{\mathbf{k}', \mathbf{k}} d_z(\mathbf{k})].
\end{align}
In terms of the critical temperatures $T_{cs, 0}$ and $T_{ct, 0}$, the Ginzburg-Landau coefficients $\alpha_1$ and $\alpha_2$ read:
\begin{align}
\alpha_1 & = N_F \int_{\mathcal{S}^2} \frac{d\Omega_{\mathbf{k}}}{4 \pi} \bigg\lbrace \mathrm{Re}\bigg[ \psi^{(0)}\bigg( \frac{1}{2} + i \frac{\lambda \| \boldsymbol{g}_{\mathbf{k}} \|}{2 \pi T} \bigg) \bigg] - \psi^{(0)} \bigg(\frac{1}{2} \bigg) \bigg\rbrace d^2_0(\mathbf{k}) - N_F \ln \bigg( \frac{T_{cs, 0}}{T} \bigg), \\
\alpha_ 2 & = N_F \int_{\mathcal{S}^2} \frac{d\Omega_{\mathbf{k}}}{4 \pi} \bigg\lbrace \mathrm{Re}\bigg[ \psi^{(0)}\bigg( \frac{1}{2} + i \frac{\lambda \| \boldsymbol{g}_{\mathbf{k}} \|}{2 \pi T} \bigg) \bigg] - \psi^{(0)} \bigg(\frac{1}{2} \bigg) \bigg\rbrace [\widehat{\boldsymbol{g}}_{\mathbf{k}} \cdot \boldsymbol{d}(\mathbf{k})]^2 - N_F \ln \bigg( \frac{T_{ct, 0}}{T} \bigg).
\end{align}
The phase difference $\theta$ between the singlet and triplet states is obtained by making the substitutions $\Delta_s = \vert \Delta_s \vert e^{i(\Phi + \theta/2)}$ and $\Delta_t = \vert \Delta_t \vert e^{i(\Phi - \theta/2)}$ into the free energy $\mathcal{F}[\Delta_s, \Delta_t]$. This yields: 
\begin{equation}\label{SM_FreeEner02}
\mathcal{F}[\Delta_s, \Delta_t] = \mathcal{F}_0 + \alpha_1 \vert \Delta_s \vert^2 + \alpha_2 \vert \Delta_t \vert^2 + \beta_1 \vert \Delta_s \vert^4 +\beta_2 \vert \Delta_t \vert^4 + 2 \gamma_1 \vert \Delta_s \vert^2 \vert \Delta_t \vert^2 + 2 \gamma_2 \cos(2\theta) \vert \Delta_s\vert^2 \vert \Delta_t \vert^2. 
\end{equation}
For completeness, we show in Fig. \ref{Gap_Amplitudes} the results of the minimization of $\mathcal{F}[\Delta_s, \Delta_t]$ for the phase transition between $d_{x^2 - y^2}$-wave and $p$-wave superconductivity inside a $B^{-}_{1g}/D_{4h}$ AM phase. We find that the $d_{x^2 - y^2}$-wave SC state exhibits coexistence within a finite range of AM interaction with the three types of $p$-wave states described in Eqs. \eqref{SM_TripletGF01}-\eqref{SM_TripletGF03}. In addition, $\gamma_2 > 0$ in the coexisting SC state, leading to $\theta = \pm \pi/2$. As a result, time-reversal symmetry becomes spontaneously broken in this case.


%